\documentclass[10pt,twoside]{article}

\usepackage{epsfig}

\def\gsim{\raise0.3ex\hbox{$>$\kern-0.75em\raise-1.1ex\hbox{$\sim$}}}
\def\lsim{\raise0.3ex\hbox{$<$\kern-0.75em\raise-1.1ex\hbox{$\sim$}}} 
\newcommand{\D}{\mbox{\rm d}}
\newcommand{\E}{\mbox{\rm e}}  
\newcommand{\beq} {\begin{equation}}
\newcommand{\eeq} {\end{equation}}
\newcommand{\psla}{p\!\!\!/}

\begin{document}

\title{
THERMODYNAMICS AND IN-MEDIUM HADRON PROPERTIES FROM LATTICE QCD
}

\author{
F. Karsch and E. Laermann \\[2pt]
{\em Fakult\"at f\"ur Physik, Universit\"at Bielefeld} \\
{\em D-33615 Bielefeld, Germany}
}
\maketitle

\thispagestyle{empty}

\begin{abstract}
Non-perturbative studies of the thermodynamics of strongly interacting 
elementary particles within the context of lattice 
regularized QCD are being reviewed.
After a short introduction into thermal QCD on the lattice
we report on the present status of investigations of bulk properties.
In particular, we discuss the present knowledge of the phase diagram
including recent developments of QCD at non-zero baryon number density.
We continue with the results obtained so far for the
transition temperature as well as the temperature dependence of energy 
and pressure and comment on screening and the heavy quark free energies. 
A major section is devoted to the discussion
of thermal modifications of hadron properties, taking special account
of recent progress through the use of the maximum entropy method.
\end{abstract}

\newpage

\tableofcontents

\newpage

\section{Introduction}

Understanding the properties of elementary particles at high temperature
and density is one of the major goals of contemporary physics. 
Through the study of properties of elementary particle matter exposed to such 
extreme conditions we hope to learn about the equation of state that controlled
the evolution of the early universe as well as the structure of compact
stars. A large experimental program is devoted to the study of hot and
dense matter created in ultrarelativistic heavy ion collisions. Lattice
studies of QCD thermodynamics have established a theoretical basis for
these experiments by providing quantitative information on the QCD 
phase transition, the equation of state and many other aspects of 
QCD thermodynamics. 

Already 20 years ago lattice calculations  
first demonstrated that a phase transition in purely
gluonic matter exists \cite{McLerran,Kuti} and that the equation
of state of gluonic matter rapidly approaches ideal gas behavior
at high temperature \cite{Bielefeld}. These
observables have been of central interest in numerical studies of the
thermodynamics of strongly interacting matter ever since. The formalism
explored in these studies, its further development and refinement has
been presented in reviews \cite{reviews} and the steady improvement
of numerical results is regularly presented at major 
conferences \cite{conferences}. 

Rather than discussing the broad spectrum of topics approached in
lattice studies of QCD thermodynamics we will concentrate here on
basic parameters, which are of direct importance for the discussion of
experimental searches for the QCD transition to the high temperature
and/or density regime, which generally is denoted as the Quark Gluon
Plasma (QGP). In our discussion of the QCD phase diagram, 
the transition temperature and the equation of state we will
also emphasize the recent progress made in lattice studies at non-zero
baryon number density. A major part of this review, however, is devoted
to a discussion of thermal modifications of hadron properties, a topic
which is of central importance for the discussion of experimental
signatures that can provide evidence for the thermal properties of the 
QGP as well as those of a dense hadronic gas.
    
\vspace{0.2cm}
\noindent
\subsection{QCD Thermodynamics}

A suitable starting point for a discussion of the equilibrium thermodynamics 
of elementary particles interacting only through the strong force is the
QCD partition function represented in terms of a  Euclidean path integral.
The grand canonical
partition function, $Z (V,T,\mu_f)$, is given as an integral over the 
fundamental quark ($\bar{\psi},\; \psi$)
and gluon ($A_\nu$) fields. In addition to its dependence
on volume ($V$), temperature ($T$) and  a set of chemical 
potentials ($\mu_f$), the partition function also depends 
on the coupling $g$ and on the quark masses
$m_f$ for $f=1,..,n_f$ different quark flavors, 

\begin{equation}
Z (V,T,\mu_f) =
\int \;{\cal D} A_\nu {\cal D}\bar{\psi}{\cal D}\psi\;
{\rm e}^{-S_E(V,T,\mu_f)} \quad .
\label{partZ}
\end{equation}
Here the bosonic fields $A_\nu$ and the Grassmann valued fermion fields
$\bar{\psi},\; \psi$ obey periodic and anti-periodic
boundary conditions in Euclidean time, respectively. The Euclidean
action $S_E\equiv S_G +S_F$ contains a purely bosonic contribution
($S_G$) expressed in terms of the field strength tensor,
$F_{\mu\nu} = \partial_\mu A_\nu - \partial_\nu A_\mu - i g  
[A_\mu,A_\nu]$, and a fermionic part ($S_F$), which couples the gluon and
quark fields through the standard minimal substitution,
\begin{eqnarray}
S_E(V,T,\mu_f) &\equiv& S_G(V,T)\; + \; S_F(V,T,\mu_f) \quad , \\
S_G(V,T)&=& \int\limits_0^{1/T} \D x_4 \int\limits_V \D^3 {\bf x} \;
\frac{1}{2} {\rm Tr}\; F_{\mu\nu} F_{\mu\nu}  \quad , \\
S_F(V,T,\mu_f)&=& \hspace{-0.1cm} \int\limits_0^{1/T}
\hspace{-0.1cm} \D x_4 \hspace{-0.1cm} 
\int\limits_V \hspace{-0.1cm} \D^3 {\bf x} \;
\sum_{f=1}^{n_f} \bar{\psi}_f \left( \gamma_\nu
    [\partial_\nu-i g A_\nu ] +\mu_f \gamma_0 + m_f \right) \psi_f .
\label{lagrangian}
\end{eqnarray} 

Basic thermodynamic quantities like the pressure ($p$) and the 
energy density ($\epsilon$) can then easily be obtained from the logarithm
of the partition function,

\begin{eqnarray}
{p\over T^4} &=& {1\over VT^3} \ln Z(T,V,\mu_f)~~, \\
{\epsilon - 3 p \over T^4} &=& T {{\rm d}\over {\rm d} T}
\left(\frac{p}{T^4} \right)_{\bigl| {\rm fixed}~\mu /T}~~.
\label{e3pmu}
\end{eqnarray}

Moreover, the phase structure of QCD can be studied by analyzing
observables which at least in certain limits are suitable order 
parameters for chiral symmetry restoration ($m_f \rightarrow 0$)
or deconfinement ($m_f \rightarrow \infty$), {\em i.e.} the chiral
condensate and its derivative the chiral susceptibility,
\begin{equation}
\langle \bar{\psi}_f \psi_f \rangle = {T\over V} {\partial \over \partial m_f} 
\ln Z(T,V,\mu_f) \quad , \quad
\chi_m = {T\over V} \sum_{f=1}^{n_f}{\partial^2 \over \partial m_f^2}
\ln Z(T,V,\mu_f) ~,
\label{chiral}
\end{equation}
as well as the expectation value of the trace of the Polyakov 
loop\footnote{A more formal definition of $\langle L \rangle$ which leads
to a well defined Polyakov loop expectation value also in the continuum limit
is given in Section 5.},
\begin{equation}
\langle L \rangle = {1\over V} \langle \sum_{\vec{x}} {\rm Tr}
L(\vec{x}) \rangle \quad ,
\label{polyakovexpect}
\end{equation}
where the trace is normalized such that ${\rm Tr} {\bf 1} = 1$.
Here $L(\vec{x})$ denotes a closed line integral over gluon fields which
represents a static quark source,  
\begin{equation}
L(\vec{x}) = {\rm e}^{-\int_0^{1/T}{\rm d}x_0 \;
A_0(x_0,\vec{x})} \quad .
\label{polyakov}
\end{equation} 
We may couple these static sources to a constant external field, $h$, and
consider its contribution to the QCD partition function. 
The corresponding susceptibility is then given by the second derivative
with respect to $h$,  
\begin{equation}
\chi_L = V\;(\langle L^2 \rangle - \langle L \rangle^2) \quad ,
\label{chiL} 
\end{equation}
where $h$ has been set to zero again after taking the derivatives.
 
\vspace{0.2cm}
\noindent
\subsection{Lattice formulation of QCD Thermodynamics}

The path integral appearing in Eq.~\ref{partZ} is regularized by
introducing
a four dimensional space-time lattice of size $N_\sigma^3 \times N_\tau$
with a lattice spacing $a$. Volume and temperature are then related
to the number of points in space and time directions, respectively,
\begin{equation}
  V= (N_\sigma\; a)^3 \quad, \quad T^{-1} = N_\tau\; a\quad ,
\end{equation}
and also chemical potentials and quark masses are expressed in units of 
the lattice spacing, $\tilde{\mu}_f = \mu_f a$, $\tilde{m}_f = m_f a$. 
The lattice spacing then does not appear explicitly as a parameter in the 
discretized version of the QCD partition function. It is controlled
through the bare couplings of the QCD Lagrangian, {\em i.e.} the gauge 
coupling\footnote{In the lattice community it is customary to introduce
instead the coupling $\beta \equiv 6/g^2$.} 
$g^2$ and quark masses $\tilde{m}_f$.   

At least on the naive level the discretization of the fermion sector
is straightforwardly achieved by replacing derivatives by
finite differences and by introducing dimensionless, Grassmann valued
fermion fields. Enforced by the requirement of gauge invariance the 
discretization of the gauge sector, however, is a bit 
more involved. Here one introduces {\it link variables} $U_\mu (x)$ which 
are associated with
the link between two neighboring sites of the lattice and describe
the parallel transport of the field $A$ from site $x$ to
the neighboring site in the $\hat{\mu}$ direction
$x+\hat{\mu}a$,
\begin{equation}
\label{link}
U_{\mu} (x) = {\rm P} \exp\biggl( i g \int_x^{x+\hat{\mu} a} \D x_\mu\,
A_\mu(x)\biggr) \quad ,
\end{equation}
where $\rm P$ denotes the path ordering. The link variables $U_\mu (x)$
are elements of the $SU(3)$ color group.

\begin{figure}[t]
\begin{center}
\epsfig{file=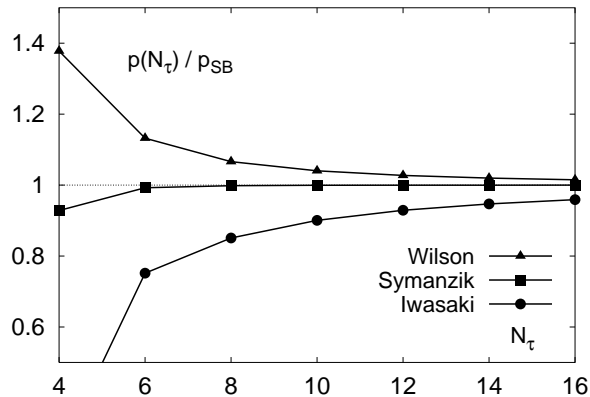,width=82mm}
\end{center}
\caption{Discretization errors in the calculation of the pressure of
a non-interacting gluon gas on lattices with temporal extent $N_\tau$.
Shown are results for the standard one plaquette action introduced by
K.~G. Wilson \protect\cite{Wilson}
and the renormalization group improved action constructed by Y.
Iwasaki \protect\cite{Iwasaki}. Both actions lead to discretization errors of
${\cal O}(a^2) \equiv {\cal O}(1/N_\tau^2)$. Also shown are results 
obtained with an Symanzik-improved action \protect\cite{Symanzik,Weisz} which
has discretization errors of ${\cal O}(a^4)$ only. 
}
\label{fig:ideal}
\end{figure}  

We will not elaborate here any further on details of the 
lattice formulation which is described in excellent 
textbooks \cite{Rothe,Muenster}. In recent years much progress has also been 
made in constructing improved discretization schemes for the 
gluonic as well as fermionic sector of the QCD Lagrangian which greatly
reduced the systematic errors introduced by the finite lattice cut-off. 
This improvement program 
and also the improvement of numerical algorithms is reviewed regularly
at lattice conferences and is discussed in  
review articles \cite{improvement}.
It has been crucial also for the calculation of thermodynamic observables
and their extrapolation to the continuum limit. The impressive accuracy that
can be achieved through a systematic analysis of finite cut-off effects
on the one hand and the use of improved actions on the other hand is 
apparent in the heavy quark mass limit of QCD, {\em i.e.} in the $SU(3)$ 
gauge theory. In particular, for bulk thermodynamic quantities
like the pressure or
energy density the discretization errors can become huge as these 
quantities depend on the fourth power of the lattice spacing $a$. 
Nonetheless, improved discretization schemes lead to a large reduction
of discretization errors and allow a safe extrapolation from lattices
with small temporal extent $N_\tau$ to the continuum limit ($N_\tau
\rightarrow \infty ,~a\rightarrow 0$) at fixed temperature $T=1/N_\tau
a$. This is illustrated in Fig.~\ref{fig:ideal} which shows results for the 
pressure, $p(N_\tau)$, in a non-interacting gluon gas calculated on a 
lattice with finite temporal extent in comparison to the 
Stefan-Boltzmann value, $p_{\rm SB}$, obtained in the continuum.  
A similarly strong reduction of cut-off effects can be achieved
in the fermion sector when using improved fermion 
actions \cite{reviews,improvement}.

\section{The QCD phase diagram}

There are many indications that strongly interacting matter at 
high temperatures/densities behaves fundamentally different
from that at low temperatures/densities. On the one hand it is 
expected that the copious production of resonances in hadronic matter 
which will occur in a hot interacting hadron gas does set a natural limit 
to hadronic physics described in terms of ordinary hadronic states. On the 
other hand it is the property of asymptotic freedom which suggests that 
the basic constituents of QCD, quarks and gluons, should propagate almost
freely at high temperature/densities. This suggests
that the non-perturbative features characterizing hadronic physics
at low energies, confinement and chiral symmetry breaking, get lost
when strongly interacting matter is heated up or compressed. 

Although the early discussions of the phase structure of
hadronic matter, e.g. based on model
equations of state, seemed to suggest that the occurrence of
a phase transition is a generic feature of strong interaction physics,
we know now that this is not at all the case. Whether the transition
from the low temperature/density hadronic regime to the high
temperature/density regime is related to a true singular
behavior of the partition function leading to a first or second order 
phase transition
or whether it is just a more or less rapid crossover phenomenon
crucially depends on the parameters of the QCD Lagrangian, {\it i.e.}
the number of light or even massless quark flavors. In particular, 
whether the transition in QCD with values of quark
masses as they are realized in nature is a true phase transition or not
is a detailed quantitative question. The answer to it most likely
is also dependent on the physical boundary conditions, {\em i.e.} 
whether the transition takes place at vanishing or non-vanishing
values of net baryon number density (chemical potential). 

At vanishing or small values
of the chemical potential the crucial control parameter is the strange
quark mass. Quite general symmetry considerations and universality
arguments for thermal phase transitions \cite{Pis84} suggest
that in the limit of light up and down quark masses the transition is 
first order if also the strange quark mass is small enough 
whereas it is just a continuous crossover for strange
quark masses larger than a certain critical value, $m^{crit}$. This
critical value also depends on the value of the light quark masses 
$m_u,~m_d$. At $m^{crit}$ the transition would be a second order phase 
transition with well defined universal properties, which are those of the 
three dimensional Ising model \cite{Gav94}.

\begin{figure}[htb]
\begin{center}
\epsfig{bbllx=0,bblly=175,bburx=564,bbury=514,
file=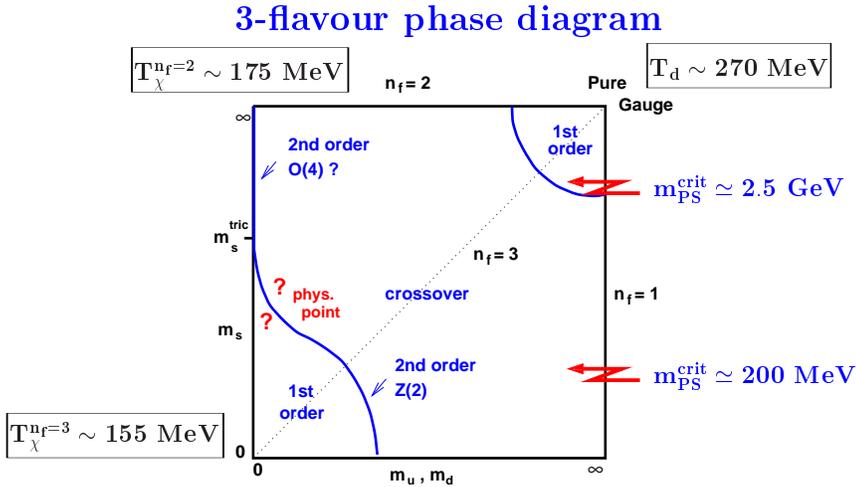,width=110mm}
\end{center}
\caption{The QCD phase diagram of 3-flavor QCD with degenerate
(u,d)-quark masses and a strange quark mass $m_s$. }
\label{fig:phased}
\end{figure}   

\begin{figure}[htb]
\begin{center}
\epsfig{file=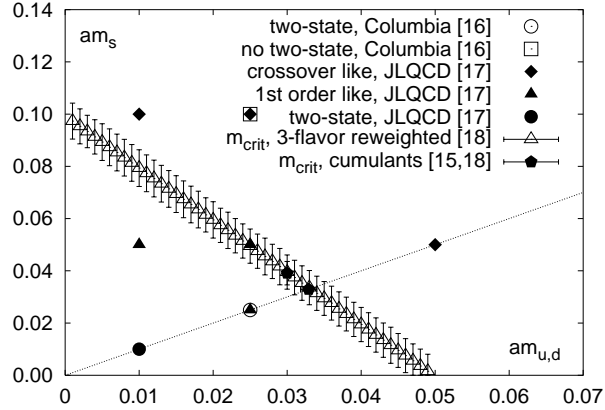,width=82mm}
\end{center}
\caption{The chiral critical line in the light and strange quark mass
plane. All the results shown have been obtained with a standard
staggered fermion discretization \protect\cite{schmidt}.
}
\label{fig:line}
\end{figure}

The dependence of the QCD phase transition on the number of flavors
and the values of the quark masses has been analyzed in quite some
detail for vanishing values of the chemical potential.
The current understanding of this phase diagram is summarized in
Fig.~\ref{fig:phased}. Its basic features
have been established in numerical calculations and are in
agreement with the general considerations based on universality
and the symmetries of the QCD Lagrangian. Nonetheless, 
the numerical values for critical temperatures and critical
masses given in this figure should just be taken as indicative; not
all of them have been determined with sufficient accuracy. 
It is, however, obvious that
there is a broad range of quark mass values or equivalently pseudo-scalar
meson masses, where the transition to the high temperature regime 
is not a phase transition but a continuous crossover. The regions of
first order transitions at large and small quark masses are separated 
from this crossover regime through lines of second order transitions
which belong to the universality class of the 3d Ising model \cite{binder}.
The {\em chiral critical line} in the small quark mass region has been
analyzed in quite some detail \cite{binder,columbia,JLQCD}. In the case
of three degenerate quark masses (3-flavor QCD) it has been verified that 
the critical point belongs to the Ising universality class \cite{binder}. 
The critical quark mass $m^{crit}$ at this point, however, is not a
universal quantity and is not yet known to a satisfying precision:
it corresponds to a pseudo-scalar mass varying between about 290~MeV
in a standard discretization and about 200~MeV in an improved one.
This result can be extended to the case of non-degenerate quark masses,
{\em i.e.} $m_s\ne m_l$ where $m_l \equiv m_u =m_d$ denotes the
value of the two light quark masses, which are taken to be degenerate.
The slope of
the chiral critical line in the vicinity of the three flavor point
can be obtained from a Taylor expansion,

\begin{equation}
m^{crit}_s = m^{crit} + 2 \;(m^{crit} - m_l) \quad .
\label{chiralline}
\end{equation} 
This relation has been
verified explicitly in a numerical calculation and is consistent with
all existing studies of the chiral transition 
line \cite{binder,columbia,JLQCD}. 
A collection of results taken from Ref.~\cite{schmidt} is shown in
Fig.~\ref{fig:line}.  

If we assume that the linear extrapolation of the chiral critical line,
Eq.~\ref{chiralline}, holds down to light quark mass values, which
correspond to the physical pion mass, we estimate that the ratio of strange
to light quark masses at this point is 
$(m^{crit}_s / m^{phys}_l)_{\mu=0} \simeq 5 - 10$, depending on the action
chosen. This clearly is too small to 
put the physical QCD point, which corresponds to $m_s / m_u \simeq 20$ into
the first order regime of the phase diagram. At vanishing chemical
potential the QCD transition thus most likely is a continuous crossover.

\begin{figure}[htb]
\begin{center}
\epsfig{file=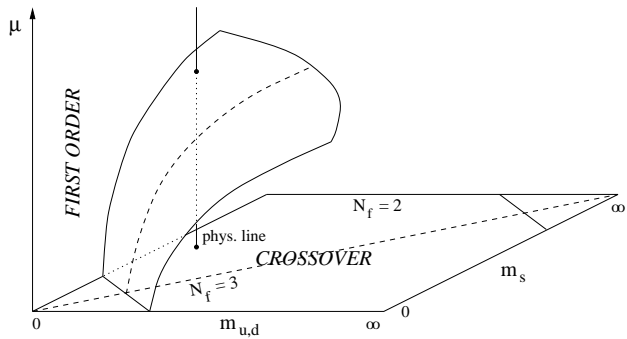,width=82mm}
\end{center}
\caption{3d sketch of the QCD phase diagram \protect\cite{schmidt}: 
Shown is the critical 
surface of second order phase transition which separates the regime 
of first order phase transitions at large values of the chemical
potential and/or small values of the light ($m_{u,d}$) and strange
($m_s$) quark masses from the regime of continuous, non-singular
transitions (crossover) to the QCD plasma phase.
}
\label{fig:phased_mue}
\end{figure}

The phase diagram shown in Fig.~\ref{fig:phased} for vanishing
quark chemical potential\footnote{The baryon chemical potential
is given by $\mu_B = 3\mu$.} ($\mu$)
can be extended to non-zero values of $\mu$. The chiral critical 
line discussed above then is part of a critical surface, $m^{crit}(\mu)$. 
For small values of the chemical potential it can be analyzed using
a Taylor expansion of the fermion determinant in terms of $(\mu / T)$.
A preliminary analysis performed for three flavor QCD \cite{schmidt} yields, 
\begin{equation}
\biggl( {m^{crit}\over T}\biggr)_\mu = \biggl( {m^{crit}\over T} \biggr)_0 
+ 0.21(6) \;  \biggl( {\mu \over T} \biggr)^2 +{\cal O}((\mu/T)^4)
\quad , \quad n_f=3~.
\label{critlinemue}
\end{equation}
The positive slope obtained for $m^{crit}(\mu)$ suggests that 
at the physical value of the strange quark mass a first order
transition can occur for values of the chemical potential larger than
a critical value determined from $m^{crit}(\mu) =m_s$
(see Fig.~\ref{fig:phased_mue}).

A first direct determination of the chiral critical point in QCD has
been performed by Fodor and Katz \cite{fodor}. As in the case of a 
Taylor expansion of the QCD partition functions in terms of the chemical
potential, which we have discussed so far, they have also performed numerical
calculations at $\mu = 0$. However, they then evaluate the exact ratios
of fermion determinants calculated at $\mu =0$ and $\mu > 0$ and use these
in the statistical reweighting of gauge field configurations\footnote{This
approach is well known under the name of Ferrenberg-Swendsen 
reweighting \cite{Ferrenberg} and finds widespread application
in statistical physics as well as in lattice QCD calculations.} to extend
the calculation to $\mu\ne 0$. They find for the chiral critical 
point \cite{fodor2}
$(\mu /T)_{crit} \simeq 1.4,~\mu^{crit}_B=3 \mu^{crit} = (725\pm 35)$~MeV.
Although this result still has to be established more 
firmly through calculations with lighter up and down quark masses
on larger lattices and with improved actions, it also
suggests that the dependence of the transition temperature on $\mu$
is rather weak. 

An alternative approach to numerical calculations at non-zero
values of the baryon number density ($\mu > 0$) is based on numerical 
calculations with imaginary 
chemical potentials \cite{laine,deForcrand,lombardo} ($\mu_I$). 
This allows straightforward numerical calculations for $\mu_I > 0$.
The results obtained in this way then have to be 
analytically continued to real values. For small values of the 
chemical potential which have been analyzed so far they turn out to 
be consistent with the results obtained with reweighting techniques.

Finally we want to note that the discussion of the dependence
of the QCD phase diagram on the baryon chemical potential in general is 
a multi-parameter problem. As pointed out in Eq.~\ref{partZ}
one generally has to deal with independent chemical potentials $\mu_f$
for each of the different quark flavors, which control the corresponding
quark number densities,
\begin{equation}
d_f = {1\over V}\; z_f \; {\partial \over \partial z_f} \ln Z(V,T,\mu_f)
\quad , \quad z_f = {\rm e}^{\mu_f /T} \quad .
\label{density}
\end{equation}
The chemical potentials thus are constrained by boundary conditions
which are enforced on the quark number densities 
by a given physical system.
In
the case of dense matter created in a heavy ion collision this is 
$\mu_s = 0$ due to the requirement that the overall strangeness content
of the system vanishes. In a dense star, on the other hand, weak decays
will lead to an equilibration of strangeness and it is the charge neutrality
of a star which controls the relative magnitude of strange and light
quark chemical potentials \cite{glendenning}.

\begin{figure}[htb]
\begin{center}
\epsfig{file=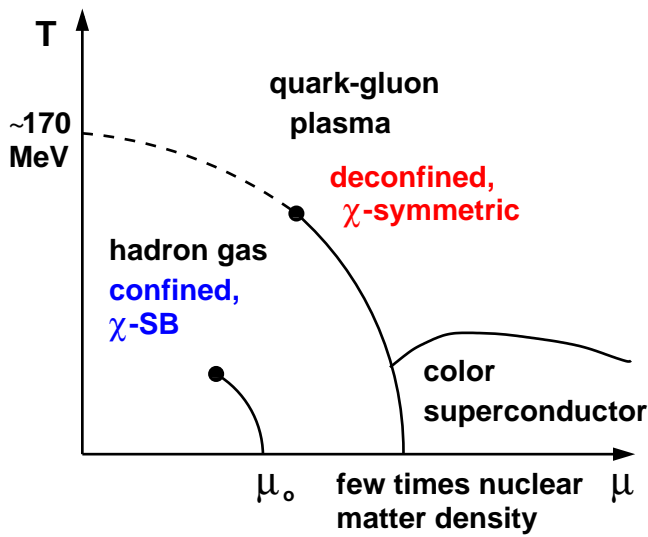,width=85mm}
\end{center}
\caption{The QCD phase diagram in the $T-\mu$ as it might look
like in the case of non-zero but light up and down quarks and a 
heavier strange quark.}  
\label{fig:phase_qgp3}
\end{figure}

So far we have discussed a particular corner of the QCD phase diagram,
{\it i.e.} the regime of small values of the chemical potential. In
fact, the numerical techniques used today to simulate QCD with
non-vanishing chemical potential seem to be reliable for $\mu /T \lsim
1$ and high temperature. Fortunately this is the regime, which also 
seems to be accessible experimentally. On the other hand there is the
entire regime of low temperature and high density which is of great
importance in the astrophysical context. In this regime it is 
expected that interesting new phases of dense matter exist. At low
temperature and asymptotically large baryon number densities asymptotic 
freedom insures that the force between quarks will be dominated by   
one-gluon exchange. As this leads to an attractive force among quarks,
it seems unavoidable
that the naive perturbative ground state is unstable at large 
baryon number density and that the formation
of a quark-quark condensate leads to a new color-superconducting phase
of cold dense matter. As this part of the phase diagram is not
in the focus of our following discussion we refer the interested reader
to the many excellent reviews which appeared in recent years \cite{css}.
In Fig.~\ref{fig:phase_qgp3} we just show a sketch of the phase diagram of
QCD for a realistic quark mass spectrum. The transition to the plasma
phase is expected to be a continuous crossover (dashed line) for small 
values of the quark chemical potential and turns into a first
order transition only beyond a critical value for the chemical
potential or baryon number density. The details of this phase diagram
in particular in the low temperature regime are, however, largely
unexplored in lattice calculations.

\section{The transition temperature}

Although a detailed analysis of the QCD phase diagram clearly is of
importance on its own, it is the thermodynamics in the region of small 
or vanishing chemical potential which is of most importance for a 
discussion of thermodynamic conditions created in heavy ion collisions
at RHIC or LHC.
Current estimates of baryon number densities obtained
at central rapidities in heavy ion collisions at RHIC \cite{pbm} 
suggest that the
baryon chemical potential is below 50~MeV, corresponding to a 
quark chemical potential of about 15~MeV. 
We will see in the following that at these small
values of $\mu$ the critical temperature is expected to
change by less than a percent from that at $\mu = 0$. 
For this reason, and of course also because much more quantitative
results are known in this case, we will in the following focus our 
discussion on the thermodynamics at $\mu = 0$.

As discussed in the previous section the transition to the high temperature
phase is continuous and non-singular for a large range of quark
masses. Nonetheless, for all quark masses the transition proceeds
rather rapidly in a small temperature interval. A definite
transition point thus can be identified, for instance through the
location of maxima of the susceptibilities of the chiral condensate, $\chi_m$ 
(Eq.~\ref{chiral}), or the Polyakov loop, $\chi_L$ (Eq.~\ref{chiL}).
While the maximum of $\chi_m$ determines the point of maximal slope 
in the chiral condensate, $\chi_L$ characterizes the change in the
long distance behavior of the heavy quark free energy 
(see section \ref{sec:poly}).

On a lattice with temporal extent $N_\tau$ and  
for a given value of the quark mass the susceptibilities define  
pseudo-critical couplings $\beta_{pc}(m_q,N_\tau)$ 
which are found to coincide within
statistical errors. In order to determine
the transition temperature $T_c=1/N_\tau a(\beta_{pc})$ one has to fix
the lattice spacing through the calculation of an experimentally or 
phenomenologically known observable. 
For instance, this can be achieved through the calculation of a hadron mass, $m_Ha$, 
or the string tension, $\sigma a^2$, at zero temperature and the same value 
of the lattice cut-off, {\it i.e.} at $\beta_{pc}$. This yields 
$T_c/\sqrt{\sigma}= 1/N_\tau \sqrt{\sigma a^2}$ and similarly for a 
hadron mass.
In the pure gauge theory the transition temperature
has been analyzed in great detail and the influence of cut-off
effects has been examined through calculations on different size
lattices and with different actions. From this one finds for the
critical temperature of the first order phase 
transition\footnote{This number is a weighted average of the
data given in Ref.~\cite{su3eos,tc_quenched}, including
a rephrasement of $r_0 T_c$ given in Ref.~\cite{Necco}
using the string model value $\alpha = \pi / 12$ for the 
$1/R$ term in the heavy quark potential. We also use $\sqrt{\sigma}= 425~$MeV
for the string tension to set the scale for $T_c$.}
\begin{eqnarray}
\underline{\rm SU(3)~gauge~theory:} && T_c/\sqrt{\sigma} = 0.632\pm
0.002 \nonumber \\
&& T_c = (269\pm 1) \; {\rm MeV}
\label{tcsu3}
\end{eqnarray}

\begin{figure}[t]
\begin{center}
\epsfig{file=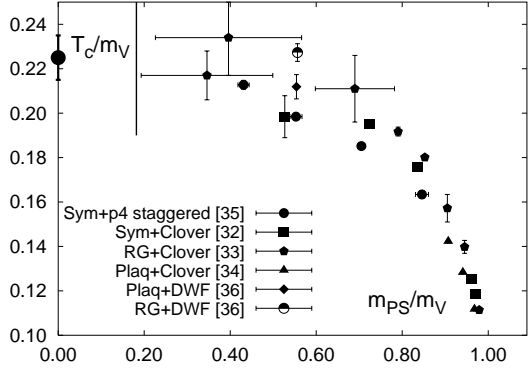,width=76mm}
\end{center}
\caption{Transition temperature in 2-flavor QCD in units of $m_V$
calculated on lattices with temporal extent $N_\tau = 4$ with different
fermion actions. 
The large dot drawn for $m_{PS}/m_V =0$
indicates the result of chiral extrapolations based on
calculations with improved Wilson \protect\cite{Ali01} as well as improved
staggered \protect\cite{Kar01} fermions.
The vertical line shows the location of the physical
limit, $m_{PS}\equiv m_\pi = 140~{\rm MeV}$.
}
\label{fig:tcnf2}
\end{figure}

Already the early calculations for the transition temperature in QCD
with dynamical quark degrees of freedom \cite{Bit91,Ber97} indicated 
that the inclusion of light quarks leads to a significant decrease of 
the transition temperature. 
A compilation of  newer results \cite{Ber97,Ali01,Edw99,Kar01,DWmass} which 
have been obtained using improved lattice regularizations for staggered
as well as Wilson fermions is presented in Fig.~\ref{fig:tcnf2}. The figure
only shows results for 2-flavor QCD obtained from calculations with 
several bare quark mass values. In order to compare calculations performed
with different actions the results are presented in terms of a {\it physical
observable}, the ratio of pseudo-scalar (pion) and vector (rho) meson
masses, $m_{PS}/m_V$. 

From Fig.~\ref{fig:tcnf2} it is evident that $T_c/m_V$ drops with increasing
ratio $m_{PS}/m_V$, {\it i.e.} with increasing quark mass. This is 
not surprising as $m_V$, of course, does not take on
the physical $\rho$-meson mass value as long as 
the quark masses used in the calculations are larger than than those
realized in nature and the ratio $m_{PS}/m_V$ thus does
not attain its physical value (vertical line in Fig.~\ref{fig:tcnf2}).
In fact, for large quark masses $m_V$ will continue to increase with $m_q$
while $T_c$ will remain finite and eventually approach the value
calculated in the pure $SU(3)$ gauge theory. The ratio
$T_c/m_V$ has to approach zero for $m_{PS}/m_V =1$.
Fig.~\ref{fig:tcnf2} alone thus does not allow us to quantify the
quark mass dependence of $T_c$.

A simple percolation picture for the QCD transition
would suggest that $T_c (m_q)$ or better $T_c(m_{PS})$ will increase
with increasing $m_q$; with increasing $m_q$
all hadrons will become heavier and it will become more and more difficult 
to excite these heavy hadronic states. It thus becomes more difficult to
create a sufficiently high particle/energy density in the
hadronic phase that can trigger a phase (percolation) transition. Such a
picture also follows from chiral model calculations \cite{chiralmodel,ch2}.
In order to check to what extent this physically well motivated picture 
finds support in the actual numerical results obtained in lattice
calculations of the quark mass dependence of the transition temperature
we should express $T_c$ in units of an observable,
which itself is not (or only weakly) dependent on $m_q$; 
the string tension (or also a hadron mass in the valence quark
chiral limit\footnote{This often is called the partially quenched limit.}) 
seems to be suitable for 
this purpose. In fact, this is what we have tacitly assumed when converting 
the critical temperature of the SU(3) gauge theory, 
$T_c/\sqrt{\sigma} \simeq 0.63$, into physical units as it has been done in 
Eq.~\ref{tcsu3}. This assumption is supported
by the observation that already in the heavy quark mass limit the
string tension calculated in units of quenched hadron masses,
e.g. $m_\rho /\sqrt{\sigma} = 1.81~(4)$~\cite{Wit97}, is in good
agreement with values required in QCD phenomenology, $\sqrt{\sigma}
\simeq 425~{\rm MeV}$.
\begin{figure}[t]
\begin{center}
\epsfig{file=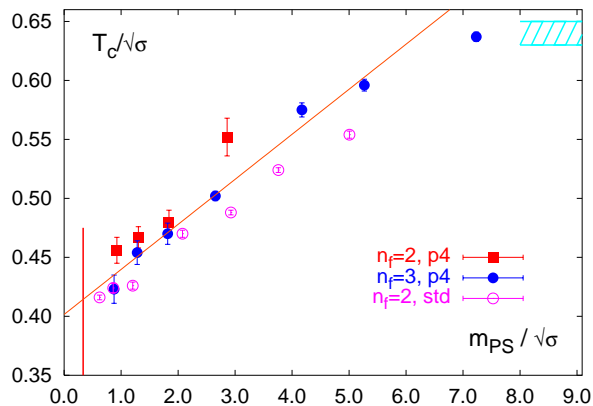,width=82mm}
\end{center}
\vskip -0.3truecm         
\caption{The transition temperature in 2 (filled squares) and 3 (circles)
flavor  QCD versus $m_{PS}/\sqrt{\sigma}$ using an improved staggered
fermion action (p4-action). Also shown are results for 2-flavor QCD
obtained with the standard staggered fermion action (open squares).
The dashed band indicates the uncertainty on $T_c/\sqrt{\sigma}$ in the
quenched limit. The straight line is the fit given in Eq.~\ref{tcfit}.}
\vskip -0.2truecm
\label{fig:tc_pion}
\end{figure}  

To quantify the quark mass dependence of the transition temperature
one may express $T_c$ in units of $\sqrt{\sigma}$.
This ratio is shown in Fig.~\ref{fig:tc_pion} as a function of
$m_{PS} / \sqrt{\sigma}$. As can be seen the transition
temperature starts deviating from the quenched value for $m_{PS}\;
\lsim \; (6-7)\sqrt{\sigma}\simeq 2.5~{\rm GeV}$. We also note that the
dependence of $T_c$ on $m_{PS}/\sqrt{\sigma}$
is almost linear in the entire mass interval.
Such a behavior might, in fact, be expected
for light quarks in the vicinity of
a $2^{nd}$ order chiral transition where the dependence of the
pseudo-critical temperature on the mass of the Goldstone-particle
follows from the scaling relation
\begin{equation}
T_c(m_{\pi}) - T_c(0) \sim m_{\pi}^{2/\beta\delta} ~.
\end{equation}
For 2-flavor QCD the critical indices $\beta$ and $\delta$ are expected
to belong to the universality class of 3-d, $O(4)$
symmetric spin models and one thus indeed would expect to find
$2/\beta\delta \simeq 1.1$.
However, this clearly cannot be the origin for the quasi linear
behavior which is observed already for rather large hadron masses and, 
moreover, seems to be independent of $n_f$.    
In fact, unlike in chiral models \cite{chiralmodel,ch2} the dependence
of $T_c$ on $m_{PS}$ turns out to be rather weak. The line shown in
Fig.~\ref{fig:tc_pion} is a fit to the 3-flavor data, 
\begin{equation}
\label{tcfit}
\biggl({T_c \over \sqrt{\sigma}} \biggr)_{m_{PS}/\sqrt{\sigma}} =
\biggl({T_c \over \sqrt{\sigma}} \biggr)_0 +
0.04(1)\; \biggl({m_{PS} \over \sqrt{\sigma}} \biggr) \quad .
\end{equation}       

It seems that the transition temperature does not react
strongly to changes of the lightest hadron masses. This favors
the interpretation that the contributions of heavy resonance masses
are equally important for the occurrence of the transition. In fact,
this also can explain why the transition
still sets in at quite low temperatures even though all hadron masses, 
including the pseudo-scalars, attain masses of the order of 1~GeV or more. 
Such an interpretation also
is consistent with the weak quark mass dependence of the critical
energy density which one finds from the analysis of the QCD equation of
state as we will discuss in the next section.

In Fig.~\ref{fig:tc_pion} we have included results from calculations
with 2 and 3 degenerate quark flavors. So far such calculations have
mainly been performed with staggered fermions.
In this case also a simulation with
non-degenerate quarks (a pair of light u,d quarks and a heavier
strange quark) has been performed. 
Unfortunately, the light quarks in this calculation are still too
heavy to represent the physical ratio of light u,d quark masses and a 
heavier strange quark mass. Nonetheless,
the results obtained so far suggest that the transition
temperature in (2+1)-flavor QCD is close to
that of 2-flavor QCD. The 3-flavor theory, on the other hand, leads to
consistently smaller values of the critical temperature,
$T_c(n_f=2)-T_c(n_f=3) \simeq 20$~MeV. Extrapolations of the transition
temperatures to the chiral limit gave
\begin{eqnarray}
\underline{\rm 2-flavor~ QCD:} &&
T_c  = \cases{
(171\pm 4)\; {\rm MeV}, & clover-improved Wilson \nonumber \cr
                       ~& fermions \cite{Ali01}   \nonumber \cr
(173\pm 8)\; {\rm MeV}, & improved staggered     \nonumber \cr
                       ~& fermions \cite{Kar01}   \nonumber \cr
             }   \nonumber \\
\underline{\rm 3-flavor~ QCD:} && 
T_c  = \; \; \;\; (154\pm 8)\; {\rm MeV}, \; \; \hspace*{0.1cm}
     \mbox{improved  staggered} \nonumber \cr
& & \hspace*{3.9cm} \mbox{fermions \cite{Kar01}}
\nonumber                    
\end{eqnarray}
Here $m_\rho$ has been used to set the scale for $T_c$.
Although the agreement between results obtained with Wilson and
staggered fermions is striking, one should bear in mind that all
these results have been obtained on lattices with temporal extent $N_\tau =4$,
{\it i.e.} at rather large lattice spacing, $a\simeq 0.3$~fm.
Moreover, there are uncertainties involved in the ansatz used to
extrapolate to the chiral limit. We thus estimate that the systematic
error on the value of $T_c /m_\rho$ still is of similar magnitude as
the purely statistical error quoted above.

\begin{figure}[t]
\vspace{9pt}
\begin{center}
\hspace*{-0.4cm}\epsfig{file=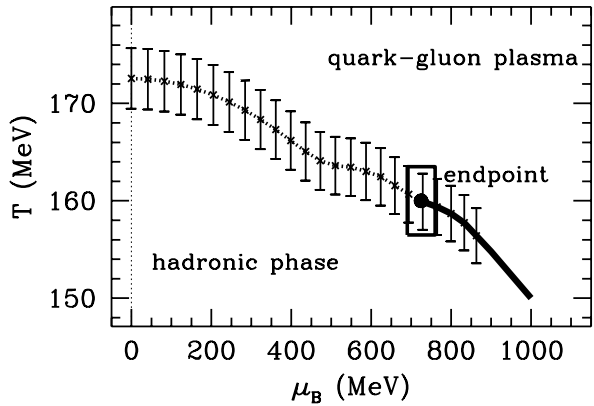,width=82mm}
\end{center}
\caption{The $\mu$-dependence of the transition temperature for
(2+1)-flavor QCD \protect\cite{fodor2}.}
\label{fig:fodor}
\end{figure}

As mentioned already in the previous section, first studies of the 
dependence of the transition temperature on the chemical potential
have been performed recently using either a statistical reweighting 
technique \cite{fodor,fodor2,Ejiri}
to extrapolate from numerical simulations performed at $\mu =0$ to
$\mu > 0$ or performing simulations with an imaginary chemical 
potential \cite{deForcrand,lombardo}
$\mu_I$ the results of which
are then analytically continued to real $\mu$. To leading
order in $\mu^2$ one finds
\begin{equation}
{T_c(\mu) \over T_c(0)} = \cases{1 - 0.0056(4) (\mu_B / T)^2 & 
                      Ref.~\cite{deForcrand} (imaginary $\mu$)
\cr
1 - 0.0078(38)(\mu_B / T)^2 & Ref.~\cite{Ejiri} (${\cal O}( \mu^2$)
reweighting)}\quad .
\label{tcmu}
\end{equation}
These results are consistent with the (2+1)-flavor calculation performed
with an exact reweighting algorithm \cite{fodor,fodor2}. The result obtained 
for $T_c(\mu)$ in this latter approach is shown in
Fig.~\ref{fig:fodor}. The dependence of $T_c$ on the chemical potential
is rather weak. We stress, however, that these calculations have not 
yet been performed with sufficiently light up and down quark masses
and a detailed analysis of the quark mass dependence has not yet been 
performed.  The $\mu$-dependence of $T_c(\mu) / T_c(0)$ is expected to
become stronger with decreasing quark masses (and, of course, vanishes in the
limit of infinite quark masses). 

\section{The equation of state}

One of the central goals in studies of the thermodynamics of
QCD is, of course, the calculation of basic thermodynamic quantities
and their temperature dependence. In particular, one wants to know the
pressure and energy density which are of fundamental
importance when discussing experimental studies of dense matter.
Besides, they allow a detailed comparison of different computational
schemes, e.g. numerical lattice calculations and analytic approaches
in the continuum.

At high temperature one generally expects that due to asymptotic 
freedom these observables show ideal gas behavior and thus are
directly proportional to the basic degrees of freedom contributing to
thermal properties of the plasma, e.g. the asymptotic behavior of the
pressure will be given by the Stefan-Boltzmann law
\begin{equation}
\lim_{T\rightarrow \infty} {p\over T^4} =  
(16 + 10.5 n_f) {\pi^2 \over 90} \quad .
\label{pressure}
\end{equation}
Perturbative calculations \cite{perteos} of corrections to this asymptotic 
behavior
are, however, badly convergent and suggest that a purely perturbative
treatment of bulk thermodynamics is trustworthy only at extremely
high temperatures, i.e. several orders of magnitude larger than the
transition temperature to the plasma phase. In analytic approaches one thus
has to go beyond perturbation theory which currently is being
attempted by either using hard thermal loop resummation techniques
in combination with a variational ansatz \cite{Bla99a,Blaizot} or
perturbative dimensional reduction combined with numerical simulations 
of the resulting effective 3-dimensional theory \cite{reduced,Laine}.  

The lattice calculation of pressure and energy density is based on the
standard thermodynamic relations given in Eq.~\ref{e3pmu}. For 
vanishing chemical potential the free energy density is directly given
by the pressure, $f=-p$.
As the partition function itself is not directly accessible in a 
Monte Carlo calculation one first takes a suitable derivative of the 
partition function, which yields a calculable expectation value, e.g.
the gauge action. After renormalizing this observable by subtracting the
zero temperature contribution it can be integrated again to obtain the 
difference of free energy densities at two temperatures,
\begin{equation}
\frac{p}{T^4} {\biggl |}_{T_o}^{T} \; = \; {1\over V} \int_{T_o}^{T}
{\rm d}t \; {\partial t^{-3} \ln Z(t,V) \over \partial t } \quad .
\label{freediv}
\end{equation} 
The lower integration limit $T_o$
is chosen at low temperatures so that $p/T_o^4$ is small and may
be ignored. This easily can be achieved in an $SU(3)$ gauge theory where
the only relevant degrees of freedom at low temperature are glueballs. 
Even the lightest ones calculated on
the lattice have large masses, $m_G \simeq 1.5$~GeV.
The free energy density thus is exponentially suppressed 
up to temperatures close
to $T_c$. In QCD with light quarks, however, the dominant contribution to the
free energy density comes from pions. 
In the small quark mass limit also $T_o$ has to be shifted to rather small
temperatures. At present, however, numerical calculations are performed
with rather heavy quarks and also the pion contribution thus is strongly
suppressed below $T_c$.

\begin{figure}[t]
\vspace{9pt}
\begin{center}
\hspace*{-0.4cm}\epsfig{file=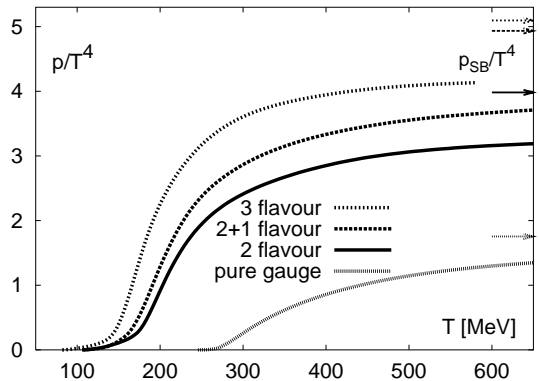,width=78mm}
\end{center}
\caption{The pressure in QCD with
two and three degenerate quark flavors as
well as two light and a heavier (strange) quark \protect\cite{Kar00a}. 
For $n_f \ne 0$
calculations have been performed on an $N_\tau=4$ lattice using
improved gauge and staggered fermion actions with a quark mass
$m/T=0.4$. Cut-off effects in these
calculations are expected to be of the order of 20\% and the pressure
is expected to become correspondingly larger once a proper extrapolation
to the continuum limit can be performed on larger lattices. At high temperature
the influence of a non-zero quark mass is expected to be small.
In the case of the SU(3)
pure gauge theory the continuum extrapolated result is shown.
}
\label{fig:qcdeosp}
\end{figure}

In Fig.~\ref{fig:qcdeosp} we show results for the pressure obtained
in calculations with different numbers of flavors \cite{Kar00a}. 
At high temperature the magnitude of $p/T^4$ clearly
reflects the change in the number of light degrees of freedom present
in the ideal gas limit. When we rescale the pressure by the corresponding
ideal gas values it becomes, however, apparent that the overall pattern
of the temperature dependence of $p/T^4$ is quite similar in all cases. 
The figure also shows that the transition region shifts to smaller 
temperatures as the number of degrees of freedom is increased.
As pointed out in the previous section such
a conclusion, of course, requires the determination of a
temperature scale that is common to all {\it QCD-like} theories which
have a particle content different from that realized in nature. We have
determined this temperature scale by assuming that the string tension is
flavor and quark mass independent. 

Other thermodynamic observables can be obtained from the pressure
using suitable derivatives. In particular one finds for the energy 
density, 
\begin{equation}
{\epsilon - 3 p \over T^4} = T {{\rm d}\over {\rm d} T}
\left(\frac{p}{T^4} \right) \quad .
\label{eands}
\end{equation}

\begin{figure}[t]
\vspace{9pt}
\begin{center}
\hspace*{-0.2cm}\epsfig{file=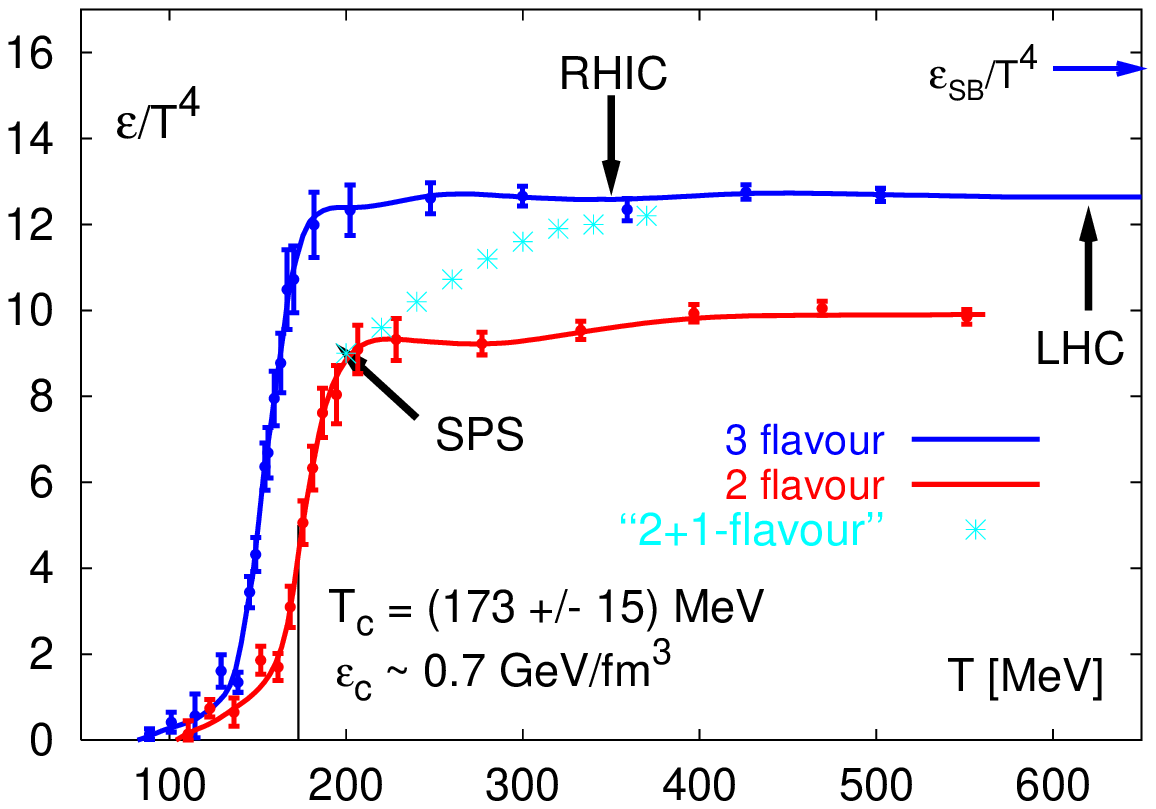,width=78mm}
\end{center}
\caption{The energy density in QCD with 2 and 3 degenerate quark flavors. 
Also shown is a sketch of the 
expected form of the energy density for QCD with a fixed strange quark
mass $m_s \sim T_c$ (see also remarks on cut-off effects in the caption
of Fig.~\ref{fig:qcdeosp}). 
The arrows indicating the energy densities reached in the initial 
stage of heavy ion collisions at the SPS, RHIC and in the future also
at the LHC are based on the Bjorken formula \protect\cite{Heinz}. 
}
\label{fig:qcdeose}
\end{figure} 

\begin{figure}[t]
\vspace{9pt}
\begin{center}
\hspace*{-0.2cm}\epsfig{file=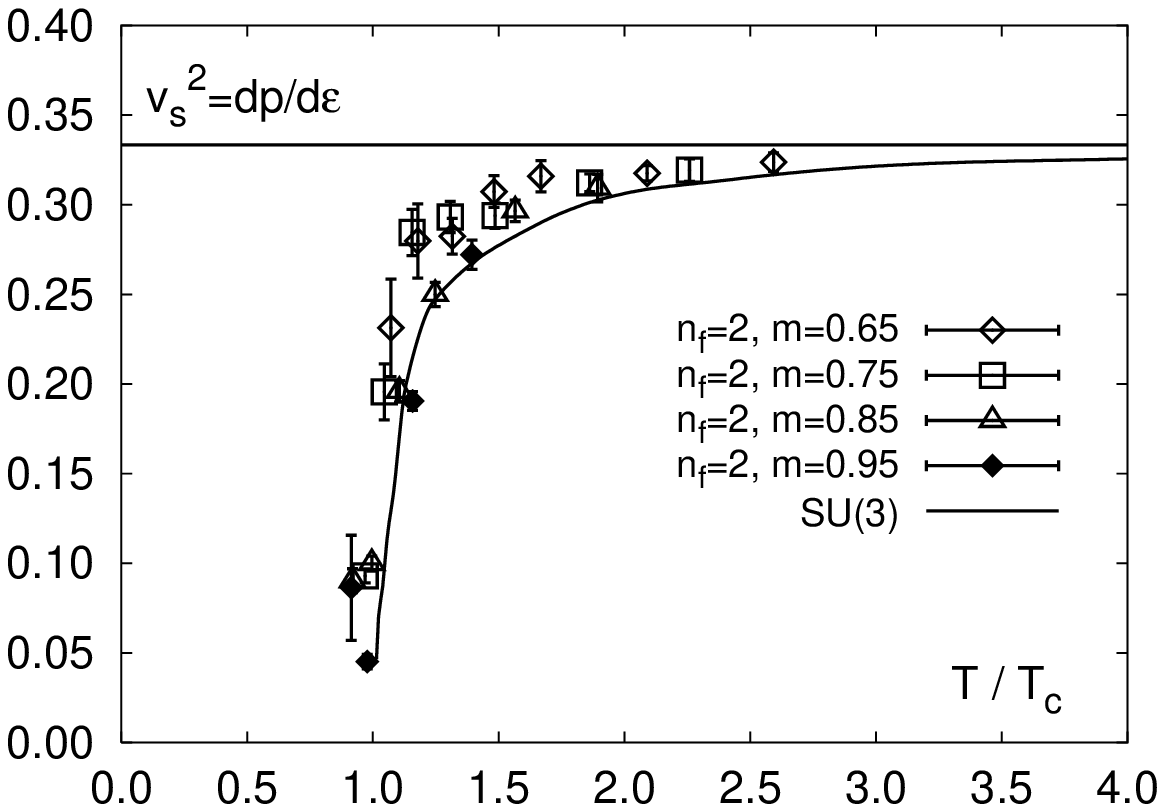,width=78mm}
\end{center}
\caption{The velocity of sound in the SU(3) gauge theory \protect\cite{su3eos} 
and in 2-flavor QCD \protect\cite{Wilson_eos}. In the latter case we show 
results from calculations with Wilson fermions performed at different values 
of the quark mass.
} 
\label{fig:sound}
\end{figure}

In Fig.~\ref{fig:qcdeose} we show results for the energy 
density\footnote{
This figure is based on data from Ref.~\cite{Kar00a} obtained for
bare quark masses $m/T=0.4$. The energy density shown does not 
contain a contribution which is proportional to the quark mass 
and thus vanishes in the chiral limit. 
}
obtained from calculations with staggered fermions and different
number of flavors. Unlike the pressure the energy density rises
rapidly at the transition temperature. Although the results shown
in this figure correspond to quark mass values in the crossover
region of the QCD phase diagram the transition clearly proceeds 
rather rapidly. This has, for instance, also consequences for
the velocity of sound, $v_s^2 = dp/d\epsilon$, which becomes rather small 
close to $T_c$.
The velocity of sound is shown in Fig.~\ref{fig:sound}. The comparison 
of results obtained from calculations in the $SU(3)$ gauge theory \cite{su3eos} 
and results obtained in simulations of 2-flavor QCD using Wilson 
fermions \cite{Wilson_eos} with different values of the quark mass
shows that the temperature dependence of $v_s$ is almost independent
of the value of the quark mass.

Also shown in Fig.~\ref{fig:qcdeose} is an estimate of the critical
energy density at which the transition to the plasma phase sets in.
In units of $T_c^4$ the transition takes place at
$\epsilon /T_c^4 \simeq 6\pm 2$ which should be compared with the 
corresponding value in the $SU(3)$ gauge theory \cite{su3eos}, 
$\epsilon /T_c^4 \simeq 0.5$. Although these numbers differ by an order
of magnitude it is rather remarkable that the transition densities
expressed in physical units are quite similar in both cases; when moving 
from large to small quark masses the increase in $\epsilon /T_c^4$ is 
compensated by the decrease in $T_c$.
This result
thus suggests that the transition to the QGP is controlled by the 
energy density, {\em i.e.} the transition seems to occur when the 
thermal system reaches a certain ``critical'' energy density. In fact,
this assumption has been used in the past to construct the phase
boundary of the QCD phase transition in the $T-\mu$ plane.

Also at non-vanishing baryon number density, the pressure as well as 
the energy density
can be calculated along the same line outlined above by using the basic
thermodynamic relation given in Eq.~\ref{e3pmu}. 
Although the statistical errors are still large, a first calculation  
of the $\mu$-dependence of the transition line indeed suggests that
$\epsilon(T_c(\mu),\mu)$ varies only little with increasing $\mu$,
$\epsilon(T_c(\mu), \mu)-\epsilon(T_c(0),0) = (1.0\pm 2.2)\mu_q^2
T_c^2(0)$~\cite{Ejiri}. First calculations of the $\mu$-dependence of 
the pressure in a wider temperature range have recently been performed
using the reweighting approach for the standard staggered fermion
formulation \cite{Fodormue} as well as the Taylor expansion for an
improved staggered fermion action up 
to {\cal O}($(\mu/T)^4$) \cite{mueeosnew}. This shows that the behavior
of bulk thermodynamic observables follow a similar pattern as in the
case of vanishing chemical potential. For instance,
the additional contribution to the pressure, $\Delta p/T^4 \equiv
(p/T^4)_{\mu/T} -(p/T^4)_{\mu=0}$ rapidly rises at $T_c$ and shows only
little temperature variation for $T/T_c \gsim 1.5$.
In this temperature regime the dominant contribution to the 
pressure arises from the contribution proportional to $(\mu /T)^2$
which also is the dominant contribution in the ideal gas limit as long as
$\mu /T \lsim 1$,
\begin{equation}
\biggl( {p \over T^4} \biggr)_{\mu /T} -
\biggl( {p \over T^4} \biggr)_{\mu =0} = {n_f \over 2}  
\biggl( {\mu \over T} \biggr)^2 +  {n_f \over 4\pi^2} 
\biggl( {\mu \over T} \biggr)^4
\quad .
\label{delmue}
\end{equation}
Here $(p/T^4)_{\mu=0}$ is given by Eq.~\ref{pressure}.
It also turns out that at non-zero values of the chemical potential
the cut-off effects in bulk thermodynamic observables are of the same
size as at $\mu =0$. Further detailed studies of the behavior of 
the pressure and the energy density thus will require a careful 
extrapolation to the continuum limit and/or the use of improved gauge
and fermion actions. 

\section{Heavy quark free energies}
\label{sec:poly}

Heavy quark free energies play a central role in our understanding
of the QCD phase transition, the properties of the plasma phase and
the temperature dependence of the heavy quark potential. The heavy
quark free energies \cite{McLerran} are defined as the QCD partition 
functions of
thermal systems containing $n~(\bar{n})$ static quark (anti-quark)
sources located at positions 
${\bf x} =\{\vec{x}_i\}_{i=1}^n$  and
$\bar{{\bf x}} =\{ \bar{\vec{x}_i} \}_{i=1}^{\bar{n}}$,
\begin{eqnarray} 
\hspace{-0.3cm}Z^{(n,\bar{n})}(V,T,{\bf x},\bar{\bf x})  
&\equiv& \exp (-F^{(n,\bar{n})} (V,T,{\bf x},\bar{\bf x})/T) 
\\
&=& \hspace{-0.1cm} 
\int \hspace{-0.1cm}{\cal D} A_\nu {\cal D}\bar{\psi}{\cal D}\psi\;
{\rm e}^{-S_E(V,T,\mu_f)}
\prod_{i=1}^n {\rm Tr}\; L(\vec{x}_i)\prod_{i=1}^{\bar{n}} 
{\rm Tr}\; L^\dagger(\bar{\vec{x}}_i)~, \nonumber
\label{Zqq}
\end{eqnarray}
where the Polyakov Loop, $L(\vec{x})$, has been defined in
Eq.~\ref{polyakov}. 
The expectation value of the product of Polyakov loops gives the
difference in free energy due to the presence of static $q\bar{q}$-sources
in a thermal heat bath of quarks and gluons,
\begin{eqnarray} 
\biggl\langle \prod_{i=1}^n {\rm Tr}\; L(\vec{x}_i)\prod_{i=1}^{\bar{n}} 
{\rm Tr}\; L^\dagger(\bar{\vec{x}}_i) 
\biggr\rangle 
&=& 
{Z^{(n,\bar{n})}(V,T,{\bf x},\bar{\bf x}) \over Z(V,T)} \\
&=& 
\exp(-\Delta F^{(n,\bar{n})} (V,T,{\bf x},\bar{\bf x})/T) \nonumber \\ 
&\equiv&
\exp(- (F^{(n,\bar{n})} (V,T,{\bf x},\bar{\bf x})-F(V,T))/T)~,
\nonumber
\label{deltaF}
\end{eqnarray}
where $Z(V,T)$ is the QCD partition function defined in Eq.~\ref{partZ}.
In particular, one considers the two point correlation function
($n=\bar{n}=1$),
\begin{equation}
G_L(|\vec{x}-\vec{y}|,T) = \langle {\rm Tr}\; L(\vec{x}) \;
{\rm Tr}\; L^\dagger (\vec{y}) 
\rangle \quad ,
\label{correlationL}
\end{equation}
and the Polyakov loop expectation value, which can be defined
through the large distance behavior of $G_L$,

\begin{equation}
\langle L\rangle = \lim_{r \rightarrow \infty}
\sqrt{G_L(r,T)} \quad , \quad
r=|\vec{x}-\vec{y}| \quad .
\label{expectationL}
\end{equation}
These observables elucidate the deconfining features of the 
transition to the high temperature phase of QCD.

\subsection{Deconfinement order parameter}

The Polyakov loop expectation value has been introduced in Section 1 as an 
order parameter for deconfinement in the heavy quark mass limit of QCD,
{\em i.e.} the $SU(3)$ gauge theory.
Like in statistical models, e.g. the Ising model, it is 
sensitive to the spontaneous breaking of a global symmetry of the 
theory under consideration. In the case of the $SU(3)$ gauge theory 
this is the global $Z(3)$ centre symmetry \cite{Yaffe}. In the presence
of light dynamical quarks this symmetry is explicitly broken and in a 
strict sense the Polyakov loop looses its property as an order
parameter. Through its relation to the two point correlation function,
Eq.~\ref{expectationL}, it however still is
the free energy of a static quark placed in a thermal heat bath, 
\begin{equation}
F_q(T) =- T \ln (\langle L\rangle ) \quad .
\label{Fq}
\end{equation}
In the low temperature, confined regime $\langle L\rangle$ is small and
the free energy thus is large. It is infinite only for an $SU(3)$ gauge
theory, {\em i.e.} for QCD in the limit $m_q \rightarrow \infty$.
In the high temperature regime however, $\langle L\rangle$ becomes large
and $F_q$ decreases rapidly when crossing the transition region.

The static quark sources introduced in the QCD partition function through
the line integral defined in Eq.~\ref{polyakov} also introduce additional
ultraviolet divergences, which require a proper renormalization. 
For the lattice regularized Polyakov loop this can be achieved through
a renormalization of the temporal gauge link variables $U_0(n_0,\vec{n})$,
{\em i.e.}
\begin{equation}
L_{\vec{n}} \equiv \prod_{i=1}^{N_\tau} Z_L(g^2) U_0(i,\vec{n}) \quad ~.
\label{Lren}
\end{equation}

The renormalization constant $Z_L(g^2)$ can, for instance, be determined by 
normalizing the two point correlation functions such that the resulting
{\em free energy of a color singlet quark anti-quark pair} at short 
distances coincides with the zero temperature heavy quark potential. 
This also insures that divergent self energy contributions
to the Polyakov loop expectation value, defined by Eq.~\ref{expectationL}, get
removed and that the heavy quark free energy can unambiguously be defined
also in the continuum limit \cite{Zantow}.

\begin{figure}[t]
\begin{center}
\epsfig{file=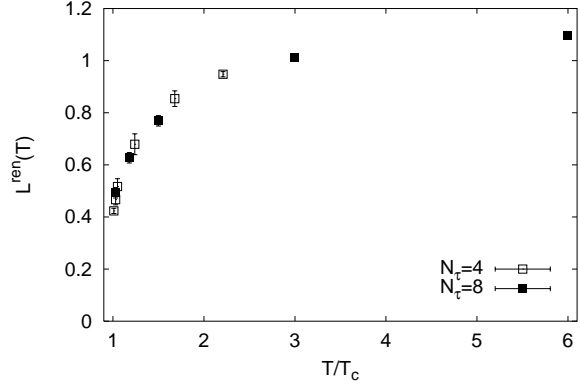,width=78mm}
\end{center}
\caption{Temperature dependence of the renormalized Polyakov loop 
expectation value in the pure $SU(3)$ gauge theory
determined from the asymptotic behavior
of color singlet free energies on lattices of size $32^3\times
N_\tau$.}
\label{fig:Lren}
\end{figure}

We stress that it is conceptually appealing to define the
renormalization constant for the Polyakov loop in terms of color
singlet free energies. Nonetheless, this can also be achieved through
the gauge invariant two point correlation functions, $G_L(r,T)$, which 
define so called color averaged free energies.
A $q\bar{q}$-pair placed in a thermal heat bath cannot maintain its
relative color orientation. The entire thermal system ($q\bar{q}$-pair
$+$ heat bath) will be colorless  and the $q\bar{q}$-pair can change
its orientation in color space when interacting with gluons of the
thermal bath. The Polyakov loop correlation function thus has to
be considered as a superposition of contributions arising from
color singlet ($F_1$) and color octet ($F_8$) contributions to the
free energy \cite{McLerran},
\begin{equation}
\E^{-F^{(1,1)}(r,T)/T } =
\frac{1}{9} \;  \E^{- F_{1}(r,T)/T } +
\frac{8}{9}   \; \E^{- F_{8}(r,T)/T }
\quad .
\label{average}
\end{equation}
At short distances the repulsive octet term is exponentially suppressed
and the contribution from the attractive singlet channel will dominate
the
heavy quark free energy,
\begin{eqnarray}
{F^{(1,1)}(r,T) \over T} &=& {F_{1}(r,T) \over T}\;  -\; \ln 9
\nonumber \\
&=& - {g^2 \over 3\pi}{1\over rT} \; -\; \ln 9
\quad\quad {\rm for}\quad \; rT << 1 \quad ,
\label{short}
\end{eqnarray}
where the last equality gives the perturbative result obtained from
1-gluon exchange at short distances. When normalizing the color singlet
free energy at short distances such that it coincides with the 
zero temperature heavy quark potential $V_{\bar{q}q}(r)$ the
corresponding color averaged free energy thus will differ by an
additive constant,
\begin{equation}
\lim_{r \rightarrow 0} (F^{(1,1)} (r,T) - F_1 (r,T)) = T\ln 9 \quad
{\rm for ~all}~T.
\label{favasym}
\end{equation}
Using this normalization condition the renormalized Polyakov loop 
order parameter has been determined for the $SU(3)$ gauge theory. It 
is shown in Fig.~\ref{fig:Lren}.
As the deconfinement phase transition is first order in the 
$SU(3)$ gauge theory the order parameter is discontinuous at $T_c$.
From the discontinuity, $\langle L \rangle (T_{c,+})\simeq 0.4$,
one finds for the change in free energy $F_\infty (T_{c,+}) \simeq
0.9 T_c \simeq 250$~MeV.

\begin{figure}[h!]
\begin{center}
\epsfig{file=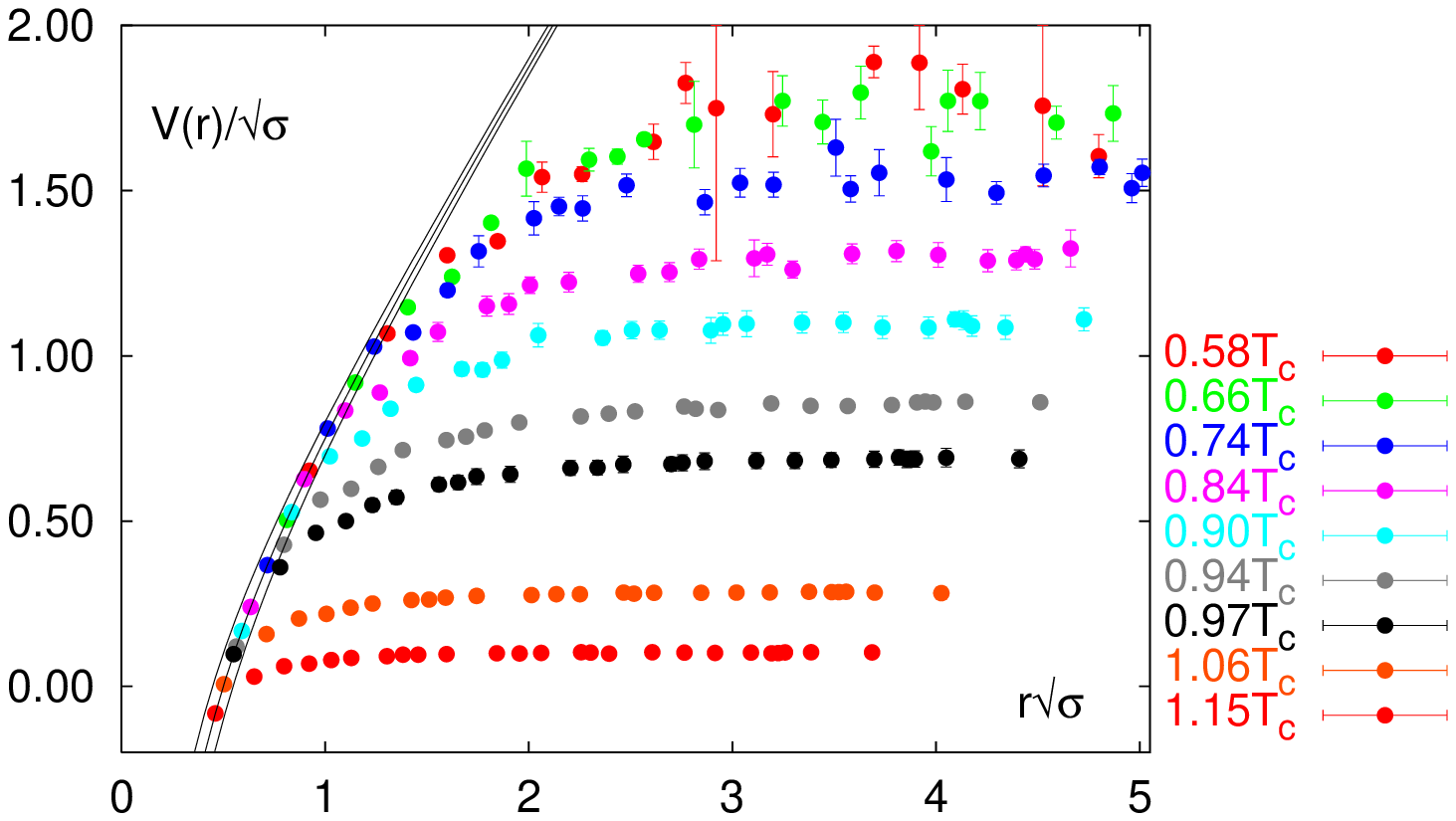,width=80mm}

\epsfig{file=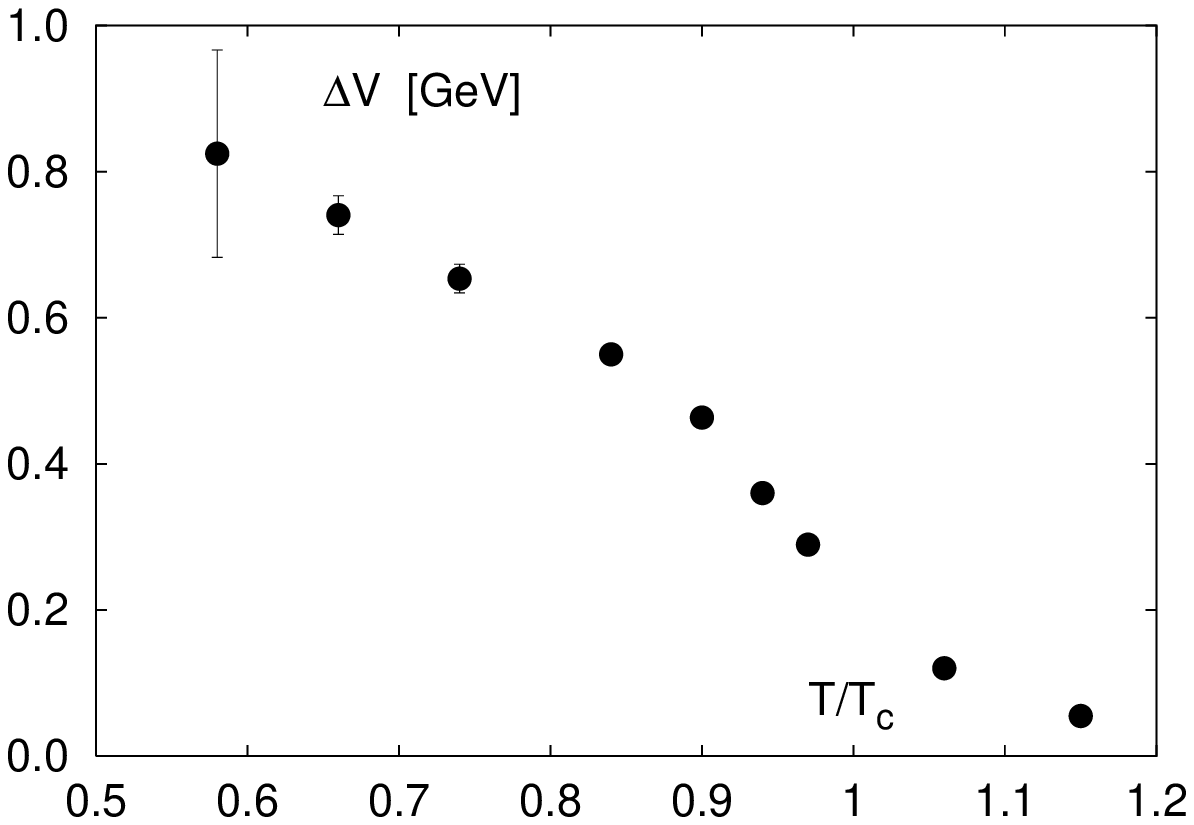,width=80mm}
\end{center}
\caption{The color averaged heavy quark free energy ($V\equiv F^{(1,1)}$)
at various temperatures 
in the low temperature phase of three flavor QCD (upper figure).
The free energy has been
normalized at the shortest distance available ($r=1/T$) to the zero
temperature Cornell potential, $V(r)/\sqrt{\sigma} = -\alpha /
r\sqrt{\sigma} + r\sqrt{\sigma}$ with $\alpha = 0.25\pm 0.05$ (solid band).
The lower figure shows the temperature dependence of the
change in free energy when separating the quark anti-quark pair from
a distance $r=0.5/\sqrt{\sigma}$ to infinity. To set the scale we
use $\sqrt{\sigma}= 425$~MeV.}
\label{fig:nf3potential}
\end{figure}

In QCD with light quarks the renormalization program outlined above has
not yet been performed in such detail as practically all studies of 
the heavy quark free energy have been performed on rather coarse lattices
with a small temporal extent, $N_\tau =4$. Nonetheless, normalizing
the free energies obtained in such calculations at the shortest distance
presently available ($rT=1/N_\tau = 0.25$) to the zero temperature
Cornell potential does seem to be a reasonable 
approximation \cite{DeTar} (see Fig.~\ref{fig:nf3potential}). Also in
this case the free energy at $T_c$ takes on a similar value as
in the pure gauge theory.

\subsection{Heavy quark potential}

The change in free energy due to the presence of a static quark anti-quark
pair is given by the two point correlation function defined in 
Eq.~\ref{correlationL}. In the zero temperature limit
the free energies, $F^{(1,1)}(r,T)$, determined from $G_L(r,T)$ define the 
heavy quark potential. Also at non zero temperature the free energies
exhibit properties expected from phenomenological discussions of 
thermal modifications of the heavy quark potential. In the pure
gauge theory $m\rightarrow \infty$) the
free energies diverge linearly at large distances
in the low temperature, confinement phase.  The 
coefficient of the linear term, the {\em string tension at finite temperature},
decreases with increasing temperature and vanishes above $T_c$. In the 
deconfined phase the free energies exhibit the behavior of a screened 
potential. At large distances they approach a constant value $F_\infty (T)$
at an exponential rate. This exponential approach defines a thermal 
screening mass. For finite quark masses the 
free energies show the expected string breaking behavior; at large
distances they approach
a finite value at all temperatures, {\it i.e.} also below $T_c$. 
This asymptotic value rapidly decreases
with decreasing quark mass and increasing temperature \cite{Kar01}
as can be seen in 
Fig.~\ref{fig:nf3potential} where we show $F^{(1,1)}(r,T)$ 
in QCD
with three light quark degrees of freedom \cite{Kar01}. 
In the lower part of Fig.~\ref{fig:nf3potential} we show
the change in free energy needed to separate a quark anti-quark
pair from a distance typical for the radius of a $c\bar{c}$ bound state,
{\it i.e.} $r=0.5/\sqrt{\sigma} \simeq 0.2$~fm, to infinity.

The change in free energy induced by a heavy quark anti-quark pair
often is taken to be {\em the heavy quark potential at finite 
temperature}. Of course, this is not quite correct and care has to
be taken when using the free energies in phenomenological discussions of
thermal properties of heavy quark bound states. In this case we 
would like to know the energy needed to break up a color singlet
state formed by a $\bar{q}q$-pair.  As pointed out in the previous subsection, 
$F^{(1,1)}$ represents a color averaged free energy. It may be decomposed in
terms of the corresponding singlet and octet contributions, which has indeed
been done for $SU(2)$ and $SU(3)$ gauge theories \cite{Philipsen,sewm}. 
However, even then one has to bear in mind that the free
energy is the difference between an energy and entropy
contributions, $F=E-TS$. Of course, from a detailed knowledge of the
temperature dependence of the free energy at fixed separation of the 
$\bar{q}q$-pair we can in principle determine
the entropy and energy contributions separately,  
\begin{equation}
S= - {\partial F(r,T) \over \partial T} \quad , \quad
E= -T^2 \; {\partial F(r,T)/T \over \partial T} \quad .
\label{ES}
\end{equation}
Such an analysis does, however, not yet exist.

\section{Thermal modifications of hadron properties}

\subsection{QCD phase transition and the hadron spectrum}

The non-perturbative structure of the QCD vacuum, 
in particular confinement
and chiral symmetry breaking, determine
many qualitative aspects of the hadron mass spectrum.
Also the actual values of light and heavy quark bound state 
masses depend on the values of the
chiral condensate and the string tension, respectively.
As these quantities change with temperature and will
change drastically close to the QCD transition temperature
it also is expected that the properties of hadrons,
e.g. their masses and widths, undergo drastic changes at
finite temperature. 

Investigations into the nature of hadronic excitations
are interesting for various reasons in the different 
temperature regimes. Below $T_c$ temperature dependent
modifications of hadron masses and widths
may lead to observable consequences in
heavy ion collision experiments e.g.
a pre-deconfined dilepton enhancement due to broadening of the
$\rho$-resonance or a shift of its mass \cite{Wambach}.
At temperatures around the
transition temperature the (approach to the)
restoration of chiral symmetry should reflect
itself in degeneracies of the hadron spectrum. First evidence
for this has, indeed, been found early in lattice calculations
of hadronic correlation functions \cite{DeTarKogut}.
In the plasma phase the very nature of hadronic excitations
is a question of interest. Asymptotic freedom leads one
to expect that the plasma consists of a gas of almost
free quarks and gluons. The lattice results on the equation of state,
however, have shown already that this is not yet the case
for the interesting temperature region.
While quasi-particle models and HTL resummed  
perturbation theory are able to reproduce the deviations from
the ideal gas behaviour observed in the
equation of state at temperatures a few times $T_c$
it remains to be seen whether
they also account properly for hadronic excitations.
This question arises in particular as the
generally assumed separation of scales
$1/T \ll 1/(gT) \ll 1/(g^2 T)$ does not hold
for temperatures quite a few times $T_c$.  

In the previous section we
have discussed modifications of the heavy quark free energy
which indicate drastic changes of the heavy quark potential in
the QCD plasma phase. As a consequence, depending on the quark
mass, heavy quark bound states cannot form above certain ``critical'' 
temperatures \cite{Matsui}. Similarly it
is expected that the QCD plasma cannot support the formation of
light quark bound states. In the pseudo-scalar sector the disappearance
of the light pions clearly is related to the vanishing of the
chiral condensate at $T_c$. For $T > T_c$ the pions would be no longer
(nearly massless) Goldstone bosons. In the plasma phase one thus may
expect to find only massive quasi-particle excitations in the
pseudo-scalar quantum number channel. However, also below $T_c$ it
is expected that the gradual disappearance of the spontaneous breaking
of chiral flavor symmetry as well as the gradual effective
restoration of the 
axial $U_A(1)$ symmetry may lead to thermal
modifications of hadron properties. 
While the  breaking of the $SU_L(n_f)\times SU_R(n_f)$ flavor symmetry
leads, for instance, to the splitting of scalar and pseudo-scalar
particle masses, the $U_A(1)$ symmetry breaking is visible in the
splitting of the pion and the $\eta^\prime$ meson.

More general, thermal modifications of the hadron spectrum should be
discussed in terms of modifications of hadronic spectral functions
which describe the thermal average over transition matrix elements
between energy eigenstates ($E_n$) with fixed quantum numbers 
($H$) \cite{LeBellac},
\begin{eqnarray}
\sigma_H(\omega, \vec p,T) =
{1 \over Z(T)} \sum_{n,m}     
e^{-E_n(\vec p)/T} (1-e^{-\omega/T})\;
\delta(\omega+E_n(\vec p)-E_m(\vec p))\cdot \nonumber\\
{|\langle n |\hat J_H(0)| m\rangle|}^2 \quad .
\label{specrepdelta}
\end{eqnarray}
These spectral functions in turn determine the structure of 
Euclidean correlation functions, $G_H(\tau,\vec{x})$. Numerical studies 
of $G_H(\tau,\vec{x})$ at finite temperature thus will allow to learn
about thermal modifications of the hadron spectrum, although in 
practice it is difficult to reconstruct the spectral functions
themselves. Here recent progress has been reached through the 
application of the maximum entropy method (MEM). Before describing these
developments and presenting recent lattice results we will start
in the next subsection with a presentation of some basic field
theoretic background on hadronic correlation functions and their
spectral representation.

\subsection{Spatial and temporal correlation functions, hadronic 
susceptibilities}
\subsubsection{Basic field theoretic background}

Numerical calculations of hadronic correlation functions are
carried out on Euclidean lattices i.e. one uses the
imaginary time formalism.
This holds also for zero temperature computations
in which case the $T \rightarrow 0$ limit has to be taken.
The formalism has been worked out in detail in 
textbooks \cite{Muenster,LeBellac,Kapusta}, however,
for readers' convenience we have collected 
some formulae in the Appendix.

Hadronic correlation functions in coordinate space,
$G_H(\tau,\vec{x})$, are defined as
\begin{eqnarray}
G_H(\tau,\vec{x}) &=&
\langle J_H (\tau, \vec{x}) J_H^{\dagger} (0, \vec{0}) \rangle
\label{eq:mesoncor}
\end{eqnarray}
where the hadronic, e.g. mesonic, currents 
$J_H (\tau,\vec{x}) =\bar{q}(\tau, \vec{x})\Gamma_H q(\tau, \vec{x})$
contain an appropriate combination of $\gamma$-matrices, $\Gamma_H$,
which fixes the quantum numbers of a meson channel; {\it i.e.,} $\Gamma_H =
1$, $\gamma_5$, $\gamma_\mu$, $\gamma_\mu \gamma_5$ for scalar,
pseudo-scalar, vector and pseudo-vector channels, respectively.

On the lattice, at zero temperature, one usually studies the temporal
correlator at fixed momentum $\vec p$. 
Save possible subtractions, the correlator is 
related to the spectral function (see Appendix),
$\sigma_H(p_0, \vec p)$, by means of
\beq
G^T_H(\tau, \vec p) =
\int_{0}^{+\infty} d p_0 \,
\sigma_H(p_0, \vec p) K(p_0,\tau)  \quad ,
\label{eq:temp_corr}
\eeq
where the kernel 
\beq
K(p_0,\tau) = 
\frac{\cosh[p_0(\tau-1/2T)]}{\sinh(p_0/2T)}
\label{eq:kernel}
\eeq
describes the propagation of a
single free boson of mass $M\equiv p_0$. 
At zero temperature, for large temporal separations the correlation
function is dominated by the exponential decay due to the
lightest contribution to the spectral function in a given channel.

At finite temperature, studies of the temporal correlator
are hampered by the limited extent of the system in the
time direction. Therefore most (lattice) analyses have
been concentrating on the spatial correlation functions.
These depend of course on the same (temperature dependent) 
spectral density but
are different Fourier transforms of it.
Projecting onto vanishing transverse momentum
and vanishing Matsubara frequency one obtains
\beq
G^S_H(z) = \int_{- \infty}^{+\infty} \frac{d p_z}{2 \pi} \, e^{i p_z z} \,
\int_{-\infty}^{+\infty} d p_0
\frac{\sigma_H(p_0, \vec 0 _\perp, p_z) }{p_0} ~~.
\label{eq:spatial0}
\eeq
In addition it is quite common in lattice calculations to analyze 
hadronic susceptibilities which are given by the space-time integral
over the Euclidean correlation functions,
\begin{eqnarray}
\chi_H 
&=& 
\int_0^{1/T} d \tau\; G^T_H(\tau, \vec 0) \quad .
\label{eq:chi}
\end{eqnarray}
The susceptibilities have a particularly simple relation to the 
spectral function,
\beq
\chi_H = 2  \int_{0}^{\infty} d p_0
\frac{\sigma_H(p_0, \vec 0)}{p_0}~~.
\label{eq:chisigma}
\eeq
Unfortunately, these susceptibilities are generally ultraviolet
divergent and the above defined integrals should be cut-off at
some short distance scale. Rather than doing this one can consider 
a closely related quantity, which provides a smooth, exponential cut-off
for the ultraviolet part of the spectral function and
is given by the thermal correlation function at $\tau T=1/2$,
\beq
G^T_H(1/2T, \vec 0) = \int_{0}^{\infty} d p_0
\frac{\sigma_H(p_0, \vec 0)}{\sinh (p_0/2T)} \quad .
\label{eq:Ghalf}
\eeq

In the case of a free stable boson of mass $M_H$ the spectral function
is a pole 
\beq
\sigma_H(p_0,\vec p) = 
\; |\langle 0 | J_H | H(\vec p) \rangle |^2 \;
\epsilon(p_0) \delta( p_0^2 - \vec p^{\,2} - M^2_H)  \quad  .
\label{eq:free_boson}
\eeq
Correspondingly, the imaginary time correlator, projected to
vanishing momentum, decreases with the mass (modulo periodicity) as
\beq
G^T_H( \tau, \vec 0) \sim \frac{1}{2 M_H}
\frac{\cosh[M_H(\tau-1/2T)]}{\sinh(M_H/2T)} \quad .
\label{eq:temp_free}
\eeq
Likewise, in this case the spatial correlator is also decaying with
the mass
\beq
G^S_H( z) \sim \frac{1}{2 M_H} \exp(-M_H z)  \quad .
\eeq
In this simple case the exponential fall-off of 
spatial and temporal correlation functions thus carry the same
information on the particle mass. Interactions in a thermal medium,
however, are likely to alter the dispersion relation to 
\beq
\omega^2(\vec p ,T) = M^2_H + {\vec p}^{\, 2} + \Pi(\vec p ,T)
\eeq
with the temperature dependent vacuum polarization
$\Pi(\vec p ,T)$.
In the simplest case, one can perhaps assume \cite{Nucu_400} that the
temperature effects can be absorbed into a temperature 
dependent mass $M_H(T)$ and a
coefficient $A(T)$ which might also be temperature dependent
and different from 1,
\beq
\omega^2(\vec p ,T) \simeq M^2_H(T) + A^2(T) {\vec p}^{\, 2}
\eeq
In this case, at zero momentum the temporal correlator will decay
with the so-called pole mass $M_H(T)$,
\beq
G^T_H(\tau,\vec 0) \sim \exp(-M_H(T) \tau)  \quad ,
\eeq
whereas the spatial correlation function has an exponential
fall-off,
\beq
G^S_H(z) \sim \exp(-M^{\rm sc}_H(T) z) \quad ,
\eeq
determined by the screening mass $M^{\rm sc}_H(T) = M_H(T)/A(T)$
which differs from the pole mass if $A(T) \neq 1$. 
In this simple case also the susceptibility defined in Eq.~\ref{eq:chi}
as well as the central value of the temporal correlation function
defined in Eq.~\ref{eq:Ghalf} are
closely related to the pole mass, 
\begin{eqnarray}
\chi_H &\sim& \frac{1}{M^2_H(T)}\quad ,  \nonumber \\
G^T_H(\tau =1/2T, \vec 0) &\sim& \frac{1}{M_H(T) \sinh (M_H(T)/2T)} \quad , 
\label{eq:sus1}
\end{eqnarray}
respectively\footnote{Note that for the spatial correlation functions the 
relevant symmetry to classify states no longer consists of the SO(3) group 
of rotations but rather
is an $SO(2) \times Z_2$ due to the asymmetry between spatial and temporal
directions. This leads to non-degeneracies between e.g. $\rho_0$ and
$\rho_{x,y}$ excitations \cite{symmetries}.}.

\begin{figure}
\begin{center}
 \epsfig{file=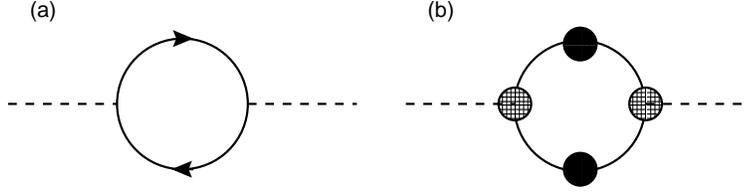,width=10.0cm}
\end{center}
\caption{The self-energy diagrams for free quarks (a) and in the
HTL approximation~(b).}
\label{fig:bubble}
\end{figure}

The opposite limit to the case of a free stable boson is 
reached for two freely propagating quarks
contributing to the spectral density. 
Here, to leading order perturbation theory the evaluation of the 
meson correlation function amounts to the 
evaluation of the self-energy diagram shown in Fig.~\ref{fig:bubble}a
in which the internal quark lines represent bare quark propagators
$S_F(i \omega_n, \vec p)$~\cite{Friman}. 

\begin{table}
\caption{Coefficients $a_H$ and $b_H$ for the free
continuum as well as coefficients $a_H^{\rm lat}$
and $d_H^{\rm lat}$ for the free lattice correlation function,
Eqs. \ref{eq:freecontinuum},\ref{eq:freelat},
in the various channels.
$d$ is given by $A/\sinh^2 E$.}

\begin{center}
{\begin{tabular}{|c|c|c|c|c|}
\hline
\rule[-2mm]{0mm}{5mm}$H$ & $a_H$ & $b_H$ & $a_H^{\rm lat}$ & $d_H^{\rm lat}$ \\
\hline
\rule[-2mm]{0mm}{5mm}$PS$                    & 1 & 0 & 1   & 0    \\
\rule[-2mm]{0mm}{4mm}$S$                     &-1 & 1 & -d  & d-1  \\
\hline
\rule[-2mm]{0mm}{6mm}$\sum_{i=1}^3 V_i$      &   &   & 3-d & d    \\
\rule[-2mm]{0mm}{5mm}$V_0$                   &   &   & 0   & -1   \\
\rule[-2mm]{0mm}{5mm}$\sum_{\mu=0}^3 V_\mu$  & 2 & 1 & 3-d & d-1  \\
\hline
\rule[-2mm]{0mm}{6mm}$\sum_{i=1}^3 AV_i$     &   &   & -2d & 2d-3 \\
\rule[-2mm]{0mm}{5mm}$AV_0$                  &   &   & 1-d & d    \\
\rule[-2mm]{0mm}{5mm}$\sum_{\mu=0}^3 AV_\mu$ &-2 & 3 & 1-3d & 3(d-1) \\
\hline
\end{tabular}}
\end{center}
\label{tab:coeff}
\end{table}  

For massless quarks the spectral density in the mesonic channel $H$ is 
then computed from Eq.~\ref{eq:freequarkprop_2} as, 
\begin{eqnarray}
\sigma_H(p_0,\vec p) = \frac{N_c}{8 \pi^2} \, (p_0^2-\vec p^{\,2})
\, a_H \,
\left\{ 
\Theta(p_0^2-\vec p^{\,2}) \,
\frac{2T}{p}
\ln\frac{\cosh(\frac{p_0+p}{4 T})}
        {\cosh(\frac{p_0-p}{4 T})} \right.\nonumber \\
 +  
\left.
\Theta(\vec p^{\,2} - p_0^2)
\left[ \frac{2T}{p}
\ln\frac{\cosh(\frac{p+p_0}{4 T})}
        {\cosh(\frac{p-p_0}{4 T})}
-\frac{p_0}{p} \right] 
\right\}
\end{eqnarray}
where $p = \sqrt{\vec p^{\,2}}$ and $a_H$ depends on the
channel analyzed (see Table~\ref{tab:coeff} for some
selected values).
In the limit of vanishing momentum $\vec p$ the
spectral density is also known \cite{Mustafa} for quarks with
non-vanishing masses $m$,
\begin{eqnarray}
\sigma_H(p_0,\vec 0) = \frac{N_c}{8 \pi^2} \, p_0^2 \,
\Theta(p_0^2- 4 m ^2) \tanh\left(\frac{p_0}{4T}\right)
\sqrt{1-\left(\frac{2m}{p_0}\right)^2} \nonumber \\
\left[a_H + \left(\frac{2m}{p_0}\right)^2 b_H \right] \quad .
\label{eq:freecontinuum}
\end{eqnarray}
The coefficients $b_H$ are also given in Table \ref{tab:coeff}.
For massless quarks closed analytic expressions
can be given for both, the temporal as well as
the spatial correlator, e.g. for the pion \cite{Friman}
\beq
G^{T, \rm free}_{\pi}(\tau, \vec 0)/T^3 = \pi N_c (1-2 \tau T)
\frac{1+\cos^2(2 \pi T \tau)}{\sin^3(2 \pi T \tau)}
+ 2 N_c \frac{\cos(2 \pi T \tau)}{\sin^2(2 \pi T \tau)}
\label{eq:free_temp}
\eeq
\begin{eqnarray}
G^{S, \rm free}_{\pi}(z) &=& 
\frac{N_c T}{4 \pi z^2 \sinh(2 \pi T z)}
[1+2 \pi T z \, \coth(2 \pi T z)] \nonumber \\
&\sim& e^{-M_{sc}^{free}z} \quad {\rm  with} \quad M_{sc}^{free} = 2\pi T 
\quad .
\label{eq:free_spat}
\end{eqnarray}
In this free field limit the susceptibility $\chi_H$, defined in
Eq.~\ref{eq:chi}, is divergent, while the temporal correlator at $\tau T =
1/2$, of course, stays finite
\beq
G^{T, \rm free}_{H}(1/2,\vec 0)/T^3 = a_H N_c/3 \quad .
\eeq
It, however, is neither related to the screening mass nor does this 
finite value have anything to do with the existence of a pole mass.

The above discussion can be extended to the leading order hard
thermal loop (HTL) approximation \cite{Mustafa} using dressed quark
propagators and vertices for the calculation of the self-energy diagrams
as indicated in Fig.~\ref{fig:bubble}b. The corresponding quark spectral
functions are given in the Appendix.

In the interacting case it has been argued \cite{Hansson}
from dimensionally reduced QCD that
the simple relation between screening mass and lowest Matsubara
frequency will be changed, to leading order, into
\beq
M_{\rm sc}(T) = 2 \pi T + \frac{C_F}{4 \pi} g^2(T) T
\left[ 2 - \ln \frac{C_F g^2(T)}{4 \pi^2} \right] \quad .
\label{eq:hansson}
\eeq
Further corrections are expected to arise from non-perturbative
effects, e.g. from the fact
that space-like Wilson loops obey an area law \cite{Borgs}. 
Correspondingly, the logarithmic potential used to arrive
at the estimate given in Eq.~\ref{eq:hansson} will turn into a linear
rising one at large distances. The effect of this 
is expected to be marginal on the screening mass, however,
will be large for so-called spatial wave functions.

\subsubsection{Lattice results}

As has been mentioned already in the previous section,
the temporal separation of the operators is limited
by the inverse temperature and the traditional approach of extracting
a ground state {\it mass} from
the long distance behavior of a Euclidean correlation function
thus is not possible. Many lattice studies
therefore have concentrated on the analysis of spatial correlation
functions. As outlined in the previous section the exponential decrease
in the spatial distance then defines a screening mass, which in certain 
limiting cases indeed may be related to the pole mass. 

\begin{figure}[t]
\begin{center}
 \epsfig{file=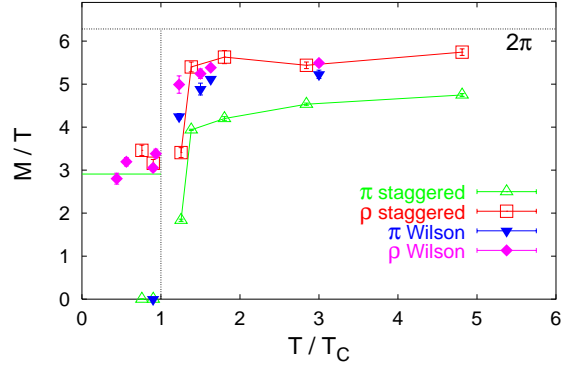,width=7.50cm}
\end{center}
\vspace{-5mm}
\caption{Screening masses in quenched QCD \protect\cite{PSchmidt,Binew} 
calculated with Wilson and staggered fermions. Below $T_c$ the masses are
normalized to $T_c$, above to $T$.
         }
\label{fig:mscreen}
\end{figure}
 
\begin{figure}
\begin{center}
 \epsfig{bbllx=100,bblly=230,bburx=460,bbury=595,clip=,
   file=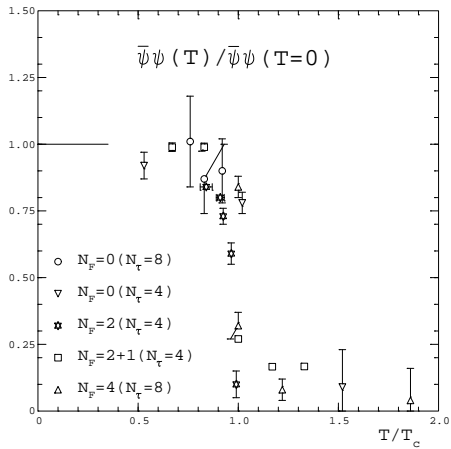,height=60mm}
\caption{The chiral condensate normalized to its zero
         temperature value for a variety of flavor numbers.
        }
\label{fig:psi}
\end{center}
\end{figure}

Fig.~\ref{fig:mscreen} shows one
representative set of lattice results on screening masses calculated
in the quenched approximation and using the Wilson as well as the staggered
discretization scheme for the fermion action.
So far, below $T_c$ the screening masses themselves have not shown any 
significant temperature dependence.
This holds for quenched staggered \cite{Sourendu} and
Wilson \cite{Nucu_400,Nucu_1,PSchmidt,Binew} as well as for 
dynamical quarks \cite{Steve2,MTc}.
Thus, there also is no substantial difference between the screening 
and the zero temperature masses. 
This seems to be in accord with sum rule predictions \cite{sumrules}
but in conflict with chiral perturbation theory 
\cite{chiralpt}.
It has been argued \cite{Brown-Rho} that the masses should 
have the same temperature
dependence as the chiral condensate. However, also the lattice results
for this quantity when normalized to its zero temperature value
do not exhibit marked deviations from unity up to temperatures
close to $T_c$, see Fig.~\ref{fig:psi}. Here it may be of particular
importance to perform numerical calculations with lighter dynamical
quark masses in order to get better control over the influence of
light virtual quarks on the chiral condensate and the resonance
decay of hadronic bound states.

At temperatures (slightly) above $T_c$ the temporal and 
spatial correlation functions of chiral partners reflect the
restoration of chiral $SU_L(n_f) \times SU_R(n_f)$ symmetry.
In particular, the vector and axial vector channel become 
degenerate \cite{DeTarKogut,Nucu_400,Nucu_1,PSchmidt,Binew,Steve2,MTc,Detar2,MILC,Boyd2,HT_ua1,gupta_lacaze,Sourendu2},
independent of the discretization and of the number of dynamical
flavors being simulated.
The same degeneracy is observed within errors in the pseudo-scalar
and isoscalar scalar ($\sigma / f_0$) channel although the latter, 
for technical reasons, 
is difficult to access on the lattice. Nevertheless, 
screening masses obtained from fits to correlation functions 
\cite{Lagae} and  
susceptibilities (at finite lattice spacing) \cite{us_crit,MILC_ua1}
lead to a consistent picture.

\begin{figure}[h!]
\begin{center}
\epsfig{file=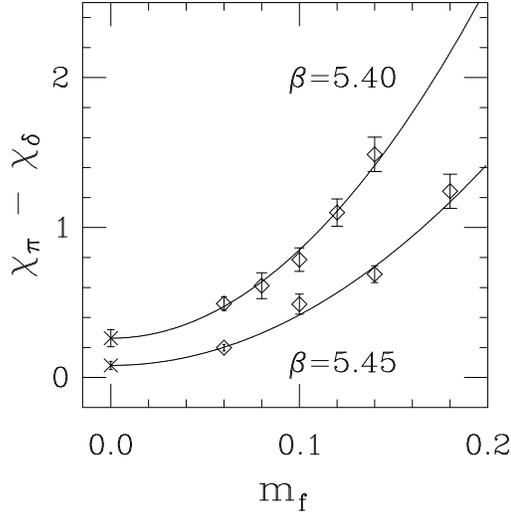,bbllx=125,bblly=385,bburx=270,bbury=530,
       width=7.00cm}
\caption{The difference of the $\pi$ and $\delta$ 
         susceptibility in two-flavor QCD as function of the 
         quark mass in lattice units.
         The data \protect\cite{Vranas} were obtained at two values 
         of the coupling constant
         which correspond to temperatures slightly above $T_c$.
         The lines are fits of the
         form $c_0 + c_2 m_f^2$ which is the quark mass dependence
         expected in the continuum. A vanishing intercept in the
         chiral limit would support effective $U_A(1)$ restoration.
        }
\label{fig:ua1}
\end{center}
\end{figure}

A more intricate question is the one concerning the degeneracy
of the pseudo-scalar and scalar isovector ($\delta/a_0$) channel
\cite{Cohen}.
In three-flavor QCD this degeneracy follows already from the above mentioned
chiral one. For two flavors in the chiral limit it
detects the effective restoration of the anomalous $U_A(1)$.
Although the $U_A(1)$ is explicitly broken by perturbative
quantum effects it might become effectively restored non-perturbatively
if topologically non-trivial zero-modes of the Dirac-matrix
are absent\footnote{In the quenched case, contributions from
   topologically non-trivial gauge configurations are not suppressed
   by powers of the quark mass arising from the fermion
   determinant \cite{Urs}.}.
If this happens already at temperatures below or at $T_c$
the chiral transition for two flavors will even be first order \cite{Pis84}.
Getting control over the zero-modes in lattice calculations is, however,
complicated. They are particularly sensitive to discretization effects 
and the continuum as well as the chiral limit have to be controlled.
The use of improved actions with better chiral properties at finite lattice 
spacing thus is important and the most convincing evidence so far has 
therefore been obtained in calculations
with the domain wall fermion discretization \cite{Vranas}
(see Fig.~\ref{fig:ua1}). 
This study suggests that the $U_A(1)$ symmetry is not yet restored 
effectively at the transition temperature, a conclusion which was 
cautiously drawn also from earlier attempts utilizing standard staggered 
actions \cite{HT_ua1,Lagae,us_crit,MILC_ua1,Christ}.

At high temperature, the screening masses are expected to approach 
the free quark propagation limit, Eq.~\ref{eq:free_spat}.
In fact, already at temperatures as low as about 1.5 $T_c$,
the results for the vector channel are close to this value,
$M_{sc} = 2 \pi T$. Nonetheless, 
Fig.~\ref{fig:mscreen} still indicates some 10 - 15 \%
deviations from free quark behavior which are of similar
size than e.g. in the equation of state. 
As such, the degeneracy between pseudo-scalar and vector channels
in the Wilson discretization \cite{Nucu_1,PSchmidt} is not
seen for staggered quarks. 
However, these results were not yet attempted to be extrapolated 
to the continuum limit and it is thus interesting that
a recent paper \cite{Sourendu2} reports that also
in the latter discretization this degeneracy is reached in the
continuum limit. 

Apart from screening masses, also the temperature dependence
of spatial wave functions has been investigated \cite{PSchmidt,MILC}.
Also in this quantity, significant differences to zero temperature
results could not be detected for temperatures $T \lsim T_c$.
Above the critical temperature, the spatial wave functions become
narrower with increasing temperature. Qualitatively, this observation
is in accord \cite{Koch} with a rising spatial 
string tension \cite{su3eos,Janos}.
However, a quantitative comparison is still lacking.

Attempts to extract masses, rather than screening masses,
from fits to temporal correlation functions have used
extended operators of various kinds possibly combined
with anisotropic lattices \cite{Nucu_400,Nucu_1,Boyd2}. 
Both methods are means to counteract the
short extent in the temperature direction.
In the first case one hopes for a ground state
contribution (if any) dominating already at small temporal
distances. In the second case the isolation of
the lowest lying state may be helped by
the increased number of data points
stabilizing more sophisticated fit ans\"atze which
include more than a ground state contribution.
However, one still has to rely on fit ans\"atze
and, if possible, use the quality of the fit to distinguish between
the various models.
Moreover, modelling the operators introduces biases.
The results have thus turned out to depend somewhat on the method.
In general, they are at least qualitatively in agreement with 
an important two-quark cut contribution.
In addition, the behavior of wave functions
was interpreted to indicate meta-stable bound states \cite{Nucu_1}.
In any case, the existence of genuine, narrow bound-states
above $T_c$  could be excluded.

\begin{figure}
\begin{center}
\epsfig{file=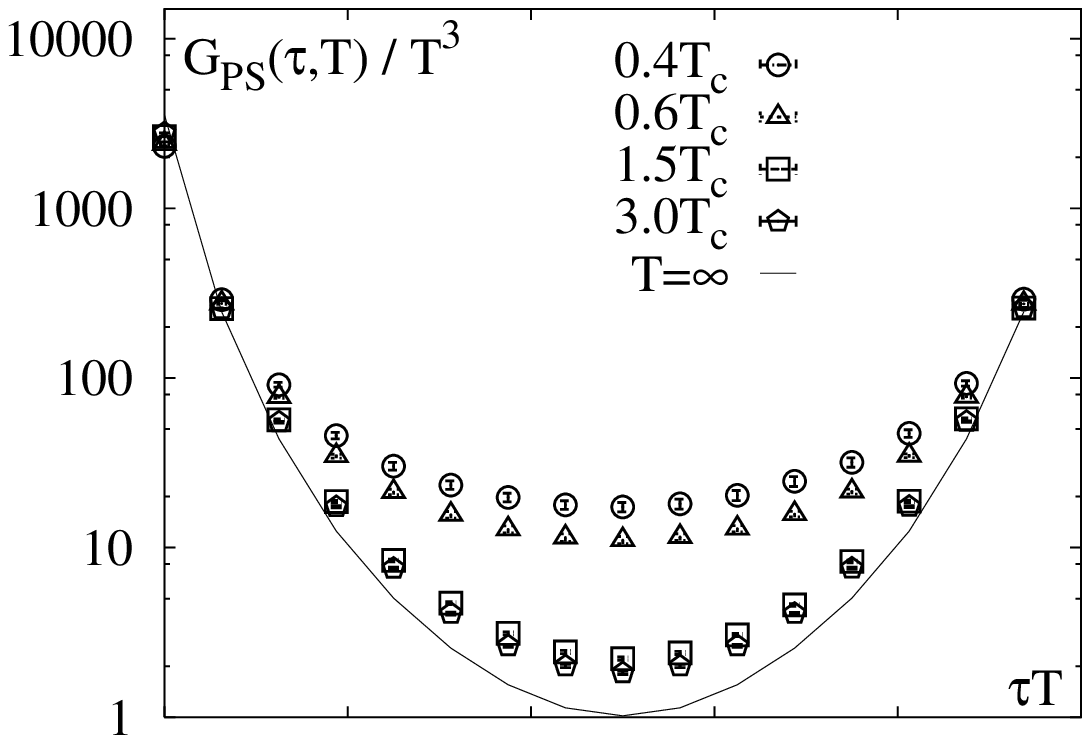,width=56mm}
\epsfig{file=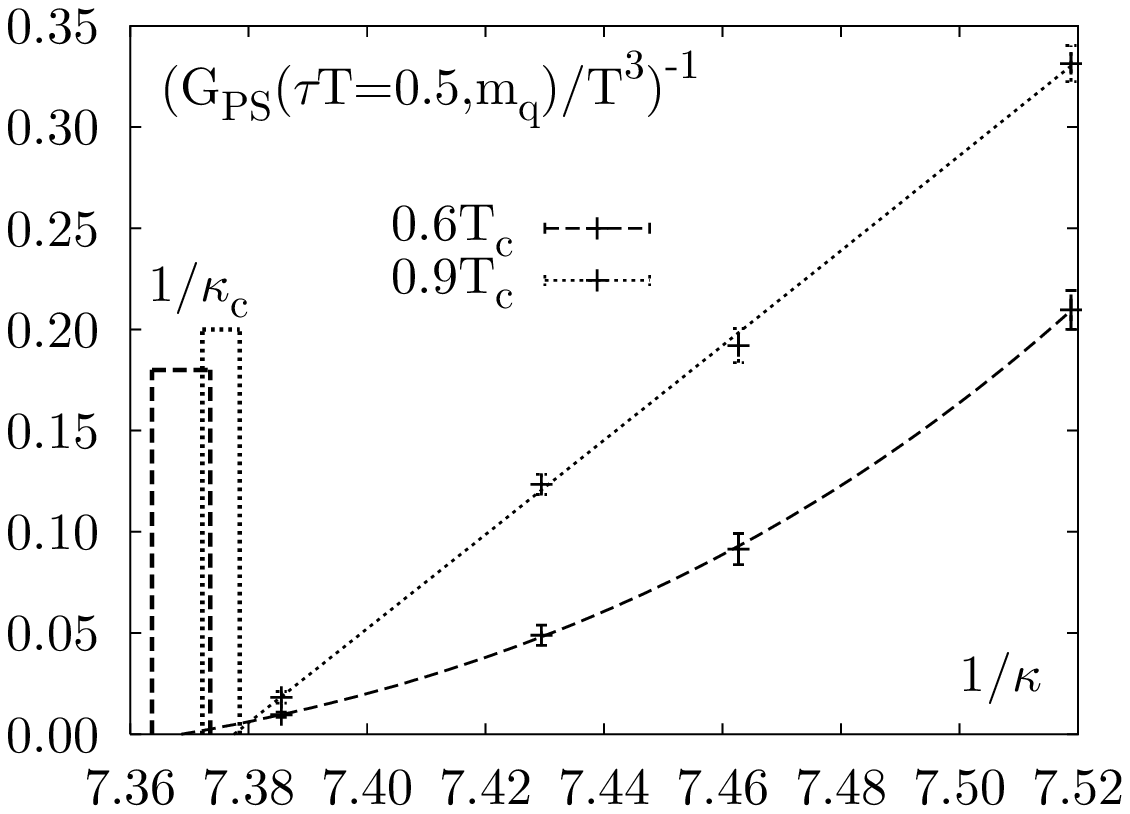,width=56mm}
\end{center}
\vspace{-3mm}
\caption{Temporal (left) pseudo-scalar correlation
functions at temperatures below and above $T_c$. Below $T_c$ the
correlation function for one value of the quark mass is shown
whereas above $T_c$ simulations have been performed in the
chiral limit. The right hand figure shows the 
chiral extrapolation of the central value of the temporal
pseudo-scalar correlator below $T_c$. The bars denote 
value and error of the critical hopping parameter obtained
from the chiral extrapolation of pseudo-scalar screening
masses.
}
\label{fig:susceptibility}
\end{figure}

\subsection{Spectral functions from hadronic correlation functions}

The discussion in the previous section shows that we have 
evidence from numerical calculations that the spectral properties
of hadrons change with temperature. The drastic changes 
of hadronic correlation functions that occur when one crosses the 
transition temperature to the high temperature plasma phase are 
self-evident from the changes in the screening masses.
The temperature dependence of correlation functions 
is particularly striking in the 
pseudo-scalar channel (see Fig.~\ref{fig:susceptibility}).
Below $T_c$ the central value, $G_{PS}(\tau T =1/2,T)$, 
diverges in the chiral limit, which 
reflects that the pseudo-scalar remains massless (Eq.~\ref{eq:sus1}),
and above $T_c$ it stays finite and comes close to the free field
limit. 
 
A much more subtle question is to quantify which thermal
modifications of the spectral functions lead to the observed
modification of the correlation functions. 
In order to decide on this we would like to have a more direct access 
to the spectral functions. It has been suggested recently \cite{Nak99} 
to apply the Maximum Entropy Method (MEM), a well known
statistical tool for the analysis of noisy data \cite{Bryan}, also
to the analysis of hadronic correlation functions. At least in
principle, {\it i.e.} given sufficiently accurate numerical 
data on large lattices, this approach allows to extract detailed information 
on hadronic spectral functions at zero as well as finite temperature.

A lattice calculation of a hadronic correlation function on a 
lattice with temporal extent $N_\tau$ provides a set of data, 
$\displaystyle{\{G_H(\tau_i)_j\}^{j=1,N_D}_{i=1,N_\tau}}$,
where $N_D$ denotes the number of measurements of a 
correlation function at a discrete set of Euclidean times
$\tau_i T = i/N_\tau,~i=1,..,N_\tau$. The statistical analysis of this
data set based on MEM aims at a determination of the {\it most likely}
spectral function which describes the data given any prior
knowledge on the correlation function or the spectral function.
This information is taken into account in the so-called default model.
In the case of hadronic correlation functions one usually tries to
provide in this way information on the perturbatively known short 
distance behavior of Euclidean correlation functions.
The first studies of spectral functions based on the 
MEM approach \cite{Nak99} and in particular the analysis of simple toy
models at zero \cite{Nak99,CPpacs02} and finite temperature \cite{Wet00} 
have, indeed, been encouraging. The MEM approach has since then been used 
and further tested in various studies of hadronic correlation 
functions at zero and finite temperature both for light and
heavy 
quarks \cite{CPpacs02,Wet01,Kar02a,Allton02,Sas02,Datta02,Asak02,Umeda02}.  
We will discuss some of the results in the following.
More details on the maximum entropy methods and its utilization
for lattice QCD calculations are given in Ref.~\cite{Asa00}.

\begin{figure}
\begin{center}
\epsfig{file=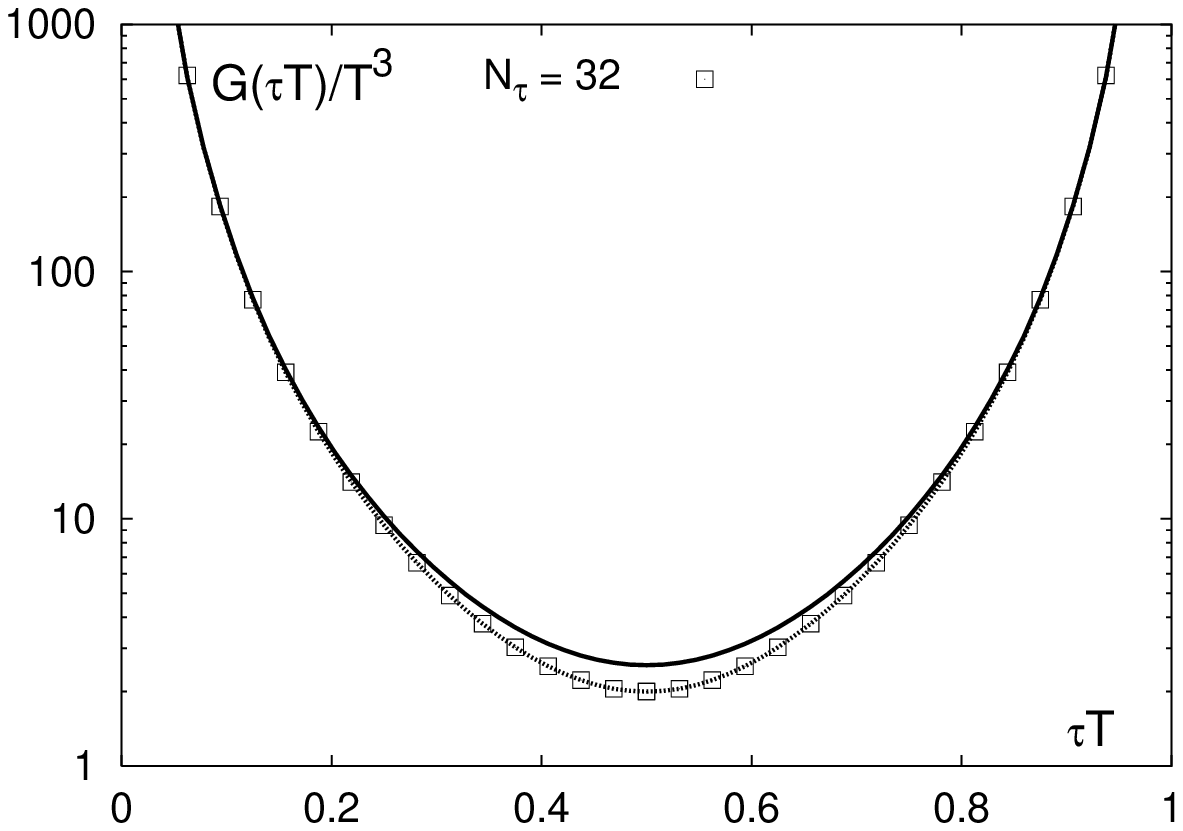,width=55mm}
\epsfig{file=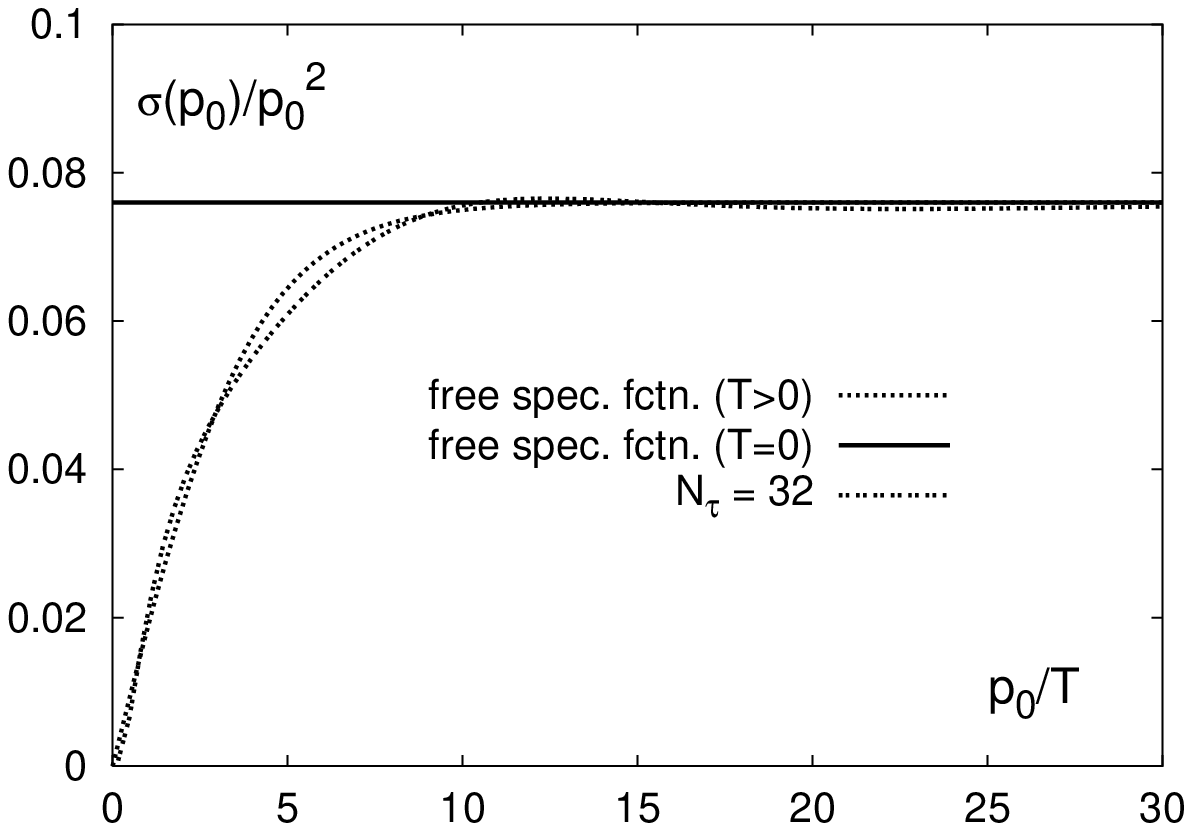,width=55mm}
\end{center}
\vspace{-3mm}
\caption{Correlation functions for a free quark anti-quark pair (left)
and the corresponding spectral functions (right). The boxes indicate 32
equally spaced points on the free correlation function for $T>0$. The 
curve through these points is obtained from the reconstructed spectral
function. The upper curve in the left hand figure and the straight line
in the right hand figure correspond to the input default model for the
MEM analysis.}
\label{fig:free}
\end{figure}

To illustrate the kind of thermal modifications of Euclidean 
correlation and spectral functions and the way this is analyzed 
using MEM let us briefly discuss some basic aspects of the 
correlation and spectral functions at zero and non-zero
temperature. To be specific we consider the vector channel.
The case of a freely propagating quark anti-quark pair is of 
relevance for the short distance part of the correlation functions
as well as in the high (infinite) temperature limit. For massless
quarks the corresponding spectral functions are given by 
Eq.~\ref{eq:freecontinuum},
\begin{equation} 
\sigma_V(p_0,\vec 0) = \cases{
\frac{N_c}{4 \pi^2} \, p_0^2 & $T=0$ \cr 
\frac{N_c}{4 \pi^2} \, p_0^2 \, \tanh\left(\frac{p_0}{4T}\right)
&$T > 0$ } \quad .
\label{eq:freeT}                        
\end{equation}
The corresponding correlation functions are shown in the left hand 
part of Fig.~\ref{fig:free}. Note that the spectral functions 
for $T=0$ as well as $T>0$ are scale invariant, {\it i.e.} the 
rescaled spectral functions, $\sigma_V /T^2$, depend on the rescaled 
energies, $p_0/T$, only. As a consequence the rescaled correlation 
functions $G_V/T^3$ are also scale invariant and are functions of 
$\tau T$ only. As a first test for the MEM analysis we may provide
a set of data corresponding to the $T>0$ correlation function shown 
in Fig.~\ref{fig:free}~(left) and use the $T=0$ spectral function
as a default model. As can be seen in Fig.~\ref{fig:free}~(right)
the MEM analysis indeed can reproduce the thermal modification
of the spectral functions already with a rather small set of 
equally spaced data points \cite{Wet00} and, in turn, yields a 
perfect reconstruction of the correlation function.

\begin{figure}
\begin{center}
\epsfig{file=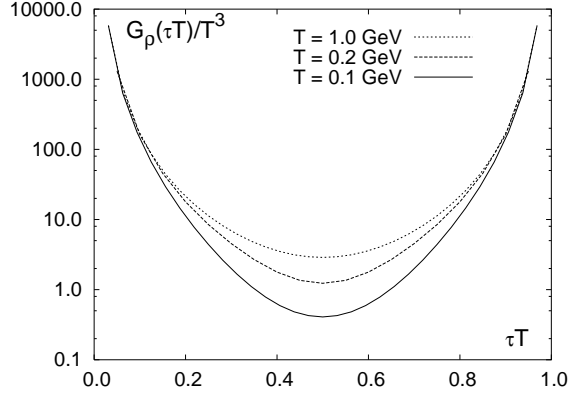,width=80mm}
\end{center}
\vspace{-4mm}
\caption{Vector meson correlation functions at temperatures
below and above $T_c$ constructed from a phenomenological zero
temperature vector meson spectral function.}
\label{fig:rhocor}
\end{figure}

At zero temperature, the free spectral function (Fig.~\ref{fig:free}), 
only characterizes the short distance behavior of the correlation function
or the large energy behavior of the spectral function, respectively.
The most prominent modification, of course, results from the 
presence of the $\rho$-resonance and a more realistic spectral function
is \cite{Shuryak},
\begin{eqnarray}
\sigma_{\rho_i} (p_0,\vec 0) &=& \frac{2p_0^2 }{\pi}
\biggl[ F_\rho^2
{\Gamma_\rho m_\rho \over (p_0^2 -m_\rho^2)^2 + 
\Gamma_\rho^2 m_\rho^2} \nonumber \\
&~&\hspace*{0.5cm} + \frac{1}{8\pi} \biggl(1 + \frac{\alpha_s}{\pi} \biggr)
{1 \over 1+{\rm exp}((\omega_0 -p_0)/\delta)} \biggr] \quad ,
\label{eq:rhospec}
\end{eqnarray}
with $\Gamma_\rho = m_\rho^3/(48\pi F_\rho^2)$ in the chiral
($m_\pi = 0$) limit and $m_\rho= 770$~MeV, $F_\rho = 142$~MeV, 
$\omega_0 = 1.3$~GeV, $\delta = 0.2$~GeV and $\alpha_s = 0.3$.
Due to the explicit appearance of hadronic scales this  
spectral function, of course, is no longer scale invariant in the sense 
introduced above. Assuming Eq.~\ref{eq:rhospec} to hold
also at non-zero temperature thus will also lead to vector 
correlation functions which are no longer scale invariant. In 
Fig.~\ref{fig:rhocor} we show the scaled vector meson correlation
functions, $G_\rho(\tau T) /T^3$, for temperatures $T=0.1,~0.2$ and 1~GeV.
The temperature dependence of the correlation function is apparent.
At the highest temperature $G_\rho (\tau T)/T^3$ almost coincides
with the corresponding rescaled free, zero temperature, correlation 
function. This shows that at these high temperatures the 
structure of the correlation function is dominated by the 
high energy part of the continuum contribution; the contribution 
of the $\rho$-resonance is suppressed.  

The {\it scaling violations} visible in Fig.~\ref{fig:rhocor} 
are most prominent for $\tau T \simeq 0.5$, which reflects that the
hadronic scales, which violate the scaling of the spectral functions
with temperature, show up at low energies. This is quite distinct from 
finite-cut off effects on the lattice that result in a distortion
of the high energy part of the spectral functions and consequently
lead to a modification of correlation functions at short distances.

\begin{figure}
\begin{center}
\epsfig{file=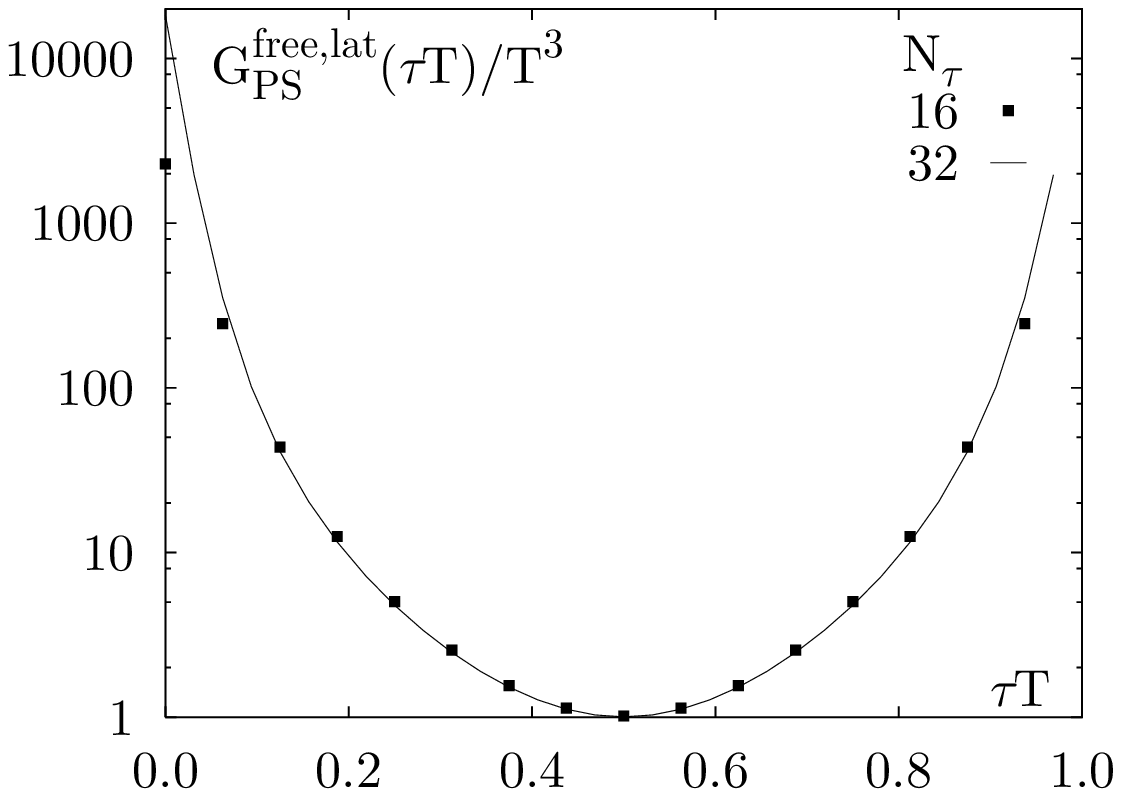,width=55mm}
\epsfig{file=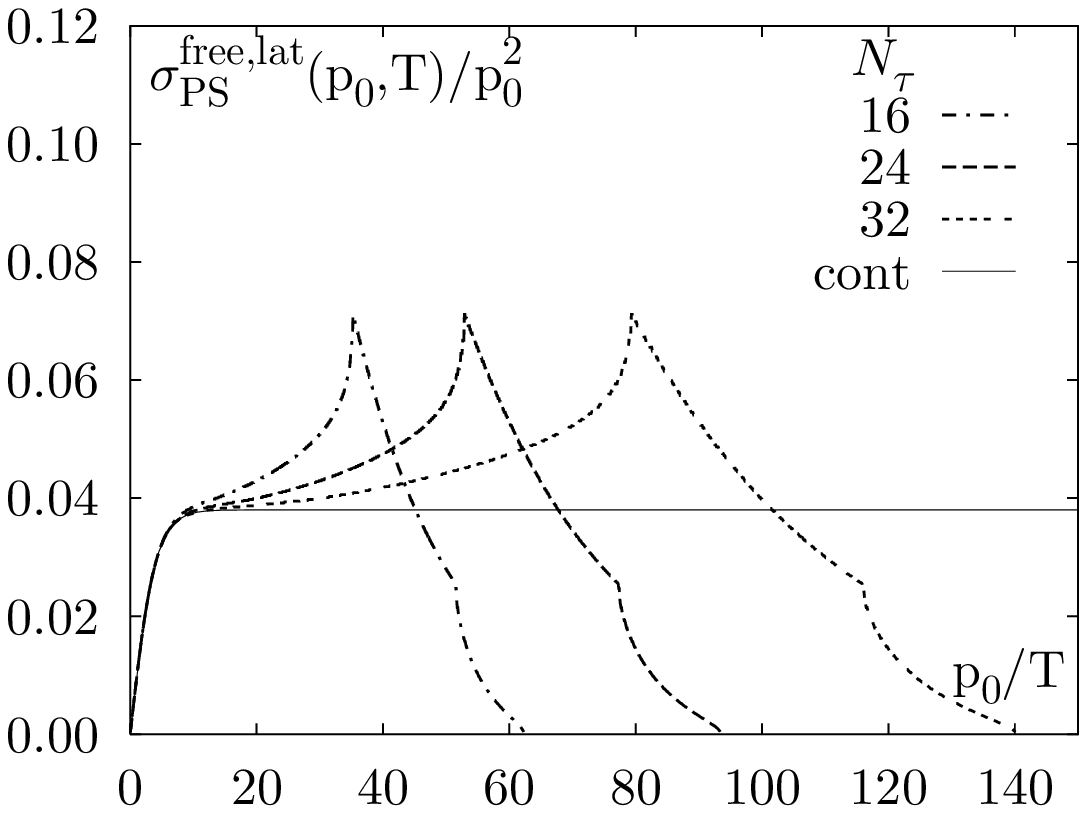,width=55mm}
\end{center}
\vspace{-3mm}
\caption{Free pseudo-scalar correlation functions (left) and 
spectral functions (right) calculated on lattices with temporal 
extent $N_\tau$ using the Wilson fermion formulation.}
\label{fig:freespec}
\end{figure}

On finite lattices the temporal correlation functions
built from free, massless quark propagators in the
Wilson discretization \cite{Carpenter} are 
given by a sum over the quark momenta $\vec k$,
at zero spatial ``meson'' momentum\footnote{
Similar formulae are available for staggered 
quarks \cite{Boyd2}.}
\begin{eqnarray}
G_H^{\rm free,lat}(\tau,\vec 0)/T^3 =
N_c \left(\frac{N_\tau}{N_\sigma}\right)^3 \sum_{\vec k}
\frac{1}{(1+B)^2} \frac{1}{\cosh^2(E N_\tau/2)} ~ \cdot \nonumber \\
\left\{ a_H^{\rm lat} \cosh[2E(\tau-N_\tau/2)] + d_H^{\rm lat} \right\} \quad .
\label{eq:freelat}
\end{eqnarray}
Here, $B$ is given by $B=2 \sum_i \sin^2 (k_i/2)$ and 
$E$ denotes the quark energy
$ E = 2 \ln \left\{ \sqrt{\alpha/4}+\sqrt{1+\alpha/4}
\right\}$
with $\alpha = (A+B^2)/(1+B)$
where $A = \sum_i \sin^2 k_i$.
The coefficients $a_H^{\rm lat}$ and $d_H^{\rm lat}$ are 
collected in Table \ref{tab:coeff}
for some channels $H$. Note that the coefficient
$d_H^{\rm lat}$ approaches 1 in the continuum limit.
Moreover, in the zero temperature limit, the $\tau$
independent piece of the correlation function
$G_H^{\rm free, lat}(\tau, \vec 0)$ vanishes
proportional to $T^3$. In Fig.~\ref{fig:freespec} we show
the free vector correlation functions and the corresponding
spectral functions \cite{freespec} on lattices with temporal 
extent $N_\tau= 16$,
24 and 32. As can be seen the influence of finite cut-off
effects is restricted to the short distance behavior of the
correlation functions, which results from the strict momentum
cut-off on a finite lattice. We also note the pronounced peak
in the lattice spectral function, which results from the distortion
of the lattice dispersion relation for free quarks at large momentum
and, in particular, close to the corners of the first Brillouin
zone. In the interacting case peaks at $a p_0 \simeq 2$ 
thus have been attributed
to the influence of the heavy Wilson doublers \cite{CPpacs02}.

\subsection{Spectral analysis of thermal correlation functions}

\begin{figure}[h]
\begin{center}
\epsfig{file=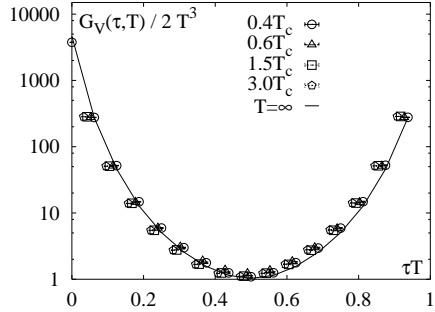,width=65mm}
\end{center}
\vspace{-3mm}
\caption{Vector meson correlation functions at various temperatures
below and above $T_c$. For better visibility data points have 
slightly been
shifted horizontally relative to the data set at $T=0.4 T_c$.
}
\label{fig:vector}
\end{figure}

In Section~6.3 we have discussed properties of the pseudo-scalar
correlation functions below and above $T_c$ which clearly suggest
that the spectral properties in this channel drastically change
when going from the low temperature hadronic phase to the high 
temperature plasma phase. After this drastic change the rescaled
pseudo-scalar correlation function above $T_c$, however, seems
to be only weakly temperature dependent.  
This property is even more pronounced in the vector
channel. In Fig.~\ref{fig:vector} we show the vector
correlation function for a set of temperatures ranging from $0.4\; T_c$
up to $3\;T_c$ calculated on lattices of size 
up to $64^3\times 16$. As can 
be seen the rescaled correlation functions
are within statistical errors temperature independent. In view
of the discussion presented in the previous section this is indeed
remarkable; the scaling of the correlation function is inconsistent
with a temperature independent spectral function and, in fact,
suggests that also the parameters determining the spectral function
(or at least its dominant part) scale with temperature.
As at high temperature the vector spectral function 
is dominated by the continuum
contribution the most obvious expectation is that the parameters
controlling the threshold in the continuum contribution, 
$\omega_0$ and $\delta$, increase with temperature. 

\begin{figure}[b]
\begin{center}
\epsfig{file=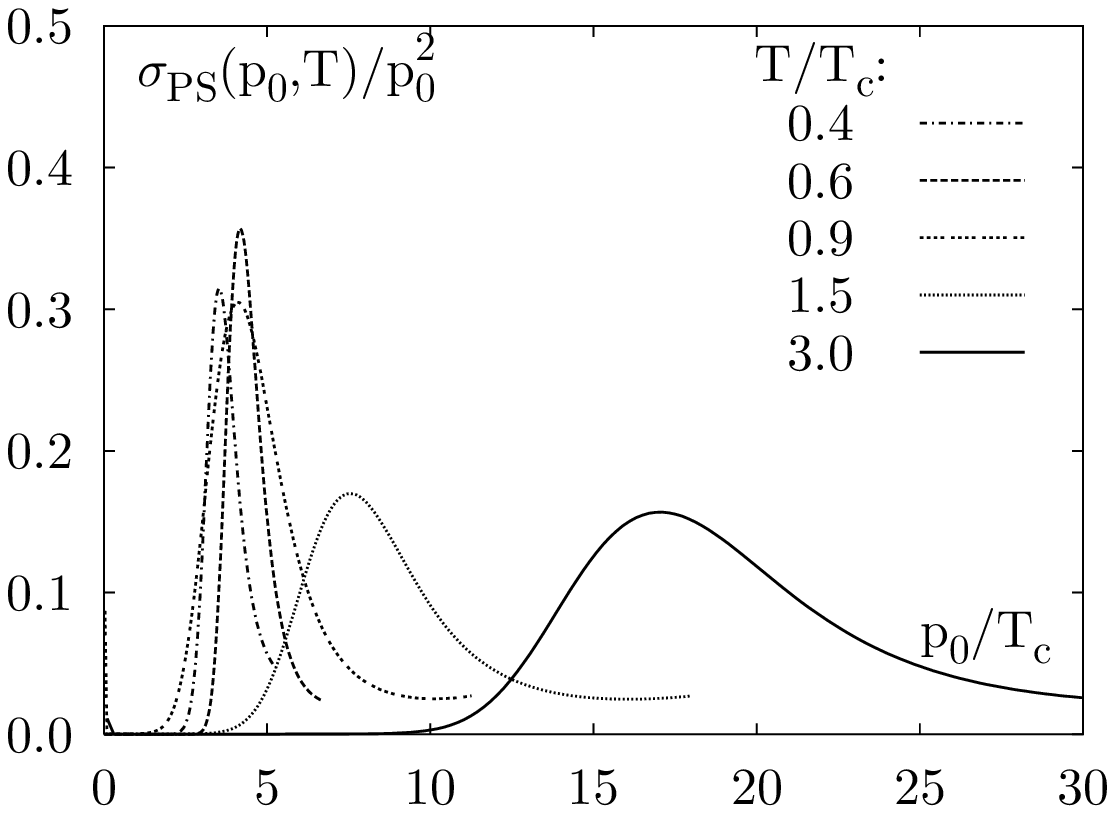,width=55mm}
\epsfig{file=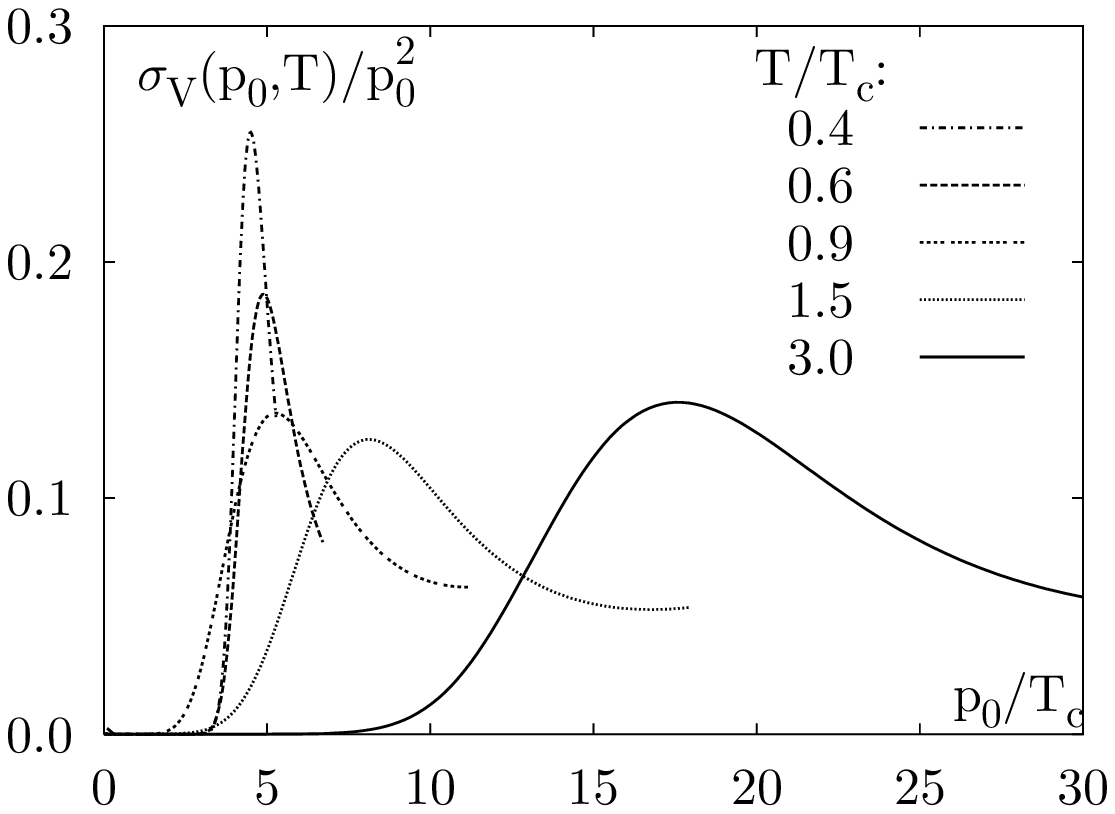,width=55mm}
\end{center}
\vspace{-3mm}
\caption{Pseudo-scalar (left) and vector (right) spectral functions obtained 
from quenched QCD calculations on lattices up to size $64^3\times 16$ using 
improved Wilson fermions.}
\label{fig:spec}
\end{figure}

The existence of a temperature dependent energy cut-off in the
vector as well as the pseudo-scalar spectral functions above $T_c$
is also obtained in the MEM analysis of the corresponding 
correlation functions. This is shown in Fig.~\ref{fig:spec}.
In both cases the reconstructed spectral functions suggest that
the contributions from the pion and rho states disappear above $T_c$.
Instead, 
a broad peak becomes visible which shifts to larger energies
with increasing temperature.

\subsection{Vector meson spectral function and thermal dilepton
rates}

The vector spectral function shown in Fig.~\ref{fig:spec} is directly
related to the thermal cross section for the production of dilepton
pairs at vanishing momentum,

\begin{equation}
{{\rm d} W \over {\rm d}p_0 {\rm d}^3p}|_{\vec{p}=0} =
{5 \alpha^2 \over 27 \pi^2} {1\over p_0^2 ({\rm e}^{p_0/T} - 1)}
\sigma_V(p_0,\vec{0},T)
\quad . 
\end{equation}  

\begin{figure}
\begin{center}
\epsfig{file=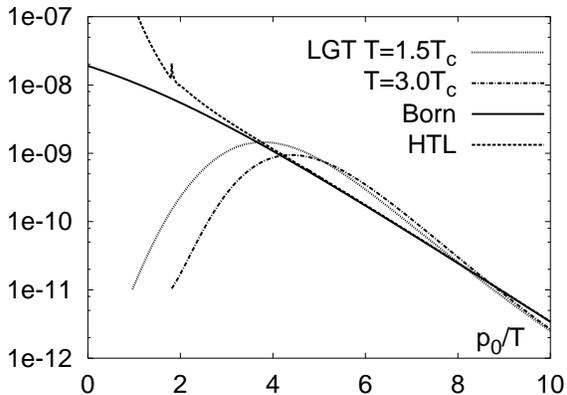,width=82mm}
\end{center}
\vspace{-3mm}
\caption{Thermal dilepton rate}
\label{fig:dilepton}
\end{figure}

This thermal dilepton rate is shown in Fig.~\ref{fig:dilepton}. The
``resonance'' like enhancement seen in Fig.~\ref{fig:spec} results
here in the enhancement of the dilepton rate over the perturbative tree 
level (Born) rate,
\begin{equation}
{{\rm d} W^{\rm Born} \over {\rm d}p_0 {\rm d}^3p}|_{\vec{p}=0} = 
{5 \alpha^2 \over 36 \pi^4} {1\over ({\rm e}^{p_0/2T} + 1)^2}
\quad ,  
\label{born}
\end{equation}
for energies $p_0 / T \in [4,8]$. Obvious discrepancies 
with hard thermal loop calcu\-la\-tions \cite{Braaten} show up at smaller energies 
where the lattice results for the spectral functions rapidly drop while the 
HTL result diverges in the infrared limit. 
In fact, this discrepancy exists already on the level of the spectral
functions themselves. The lattice spectral functions vanish (rapidly)
in the limit $p_0 \rightarrow 0$ whereas in the vector channel 
the HTL-spectral function diverges \cite{Mustafa}. As we have pointed out 
in the discussion of the $\rho$-meson spectral function at $T=0$, 
with increasing temperature it becomes more and more difficult to resolve 
the low energy part of spectral functions. The suppression of the 
thermal dilepton rate at low energies deduced from the structure
of $\sigma_V$ thus will require further detailed investigations
on large lattices. However, even without invoking the MEM analysis
to determine $\sigma_V(p_0,T)$ one can deduce a minimal constraint on
the low energy behavior of the spectral function from the fact that
the correlation function $G_V(\tau,\vec{0})$ stays finite at high
temperature. This demands that $\sigma_V(p_0,T) \sim p_0^a$ with 
$a > 0$.  The vector spectral function thus has to vanish
in the limit $p_0 \rightarrow  0$. In fact, in order to get
non-vanishing transport coefficients in the QGP the spectral
function should be proportional to $p_0$ in this limit \cite{Aarts}.
This may be included in a MEM analysis as an additional
constraint \cite{Gupta}. In any case,
getting the low energy behavior of the vector spectral functions
under control still is a challenge for future studies.

\subsection{Heavy quark spectral functions and charmonium suppression}

As the properties of mesons constructed from light quarks are closely
related to chiral properties of QCD it is expected that these states
dissolve in the QGP. The situation, however, is different for heavy quark
bound states, which generally are expected to be controlled by
properties of the heavy quark potential at moderate distances ($R\lsim
0.5fm$). Moreover, this distance scale will depend on the quark flavor
under consideration (charmonium, bottonium). Heavy quark bound states
thus are sensitive to screening of the heavy quark potential at high
temperature and therefore probe the
deconfining aspect of the QCD phase transition \cite{Matsui}. Whether
a heavy quark bound state survives the QCD phase transition
or not strongly depends on the efficient screening of the interaction
among quarks and anti-quarks in the QGP. Potential model calculations
indicate that some $c\bar{c}$ bound states, e.g. the $J/\psi$ and 
in particular the $b\bar{b}$ bound states
could survive at temperatures close to $T_c$ while radially excited
states like the $\chi_c$ most likely get dissolved at $T_c$~\cite{Satz}.

\begin{figure}
\begin{center}
\epsfig{file=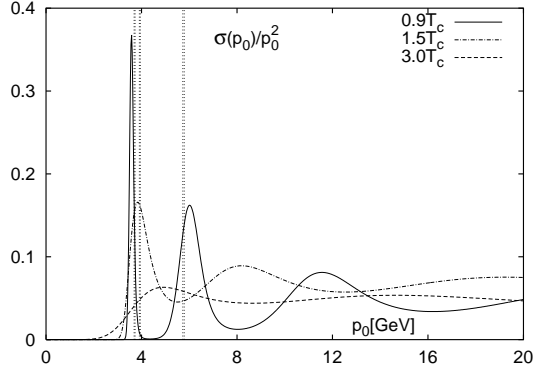,width=75mm}
\end{center}
\caption{Heavy quark spectral functions and screening masses (vertical
lines) in the vector channel calculated in quenched QCD on lattices of 
size $40^4$ ($T\simeq 0.9 T_c$), $64^3 \times 24$ ($T\simeq 1.5 T_c$)
and $48^3\times 12$ ($T\simeq 3 T_c$) at $\beta=7.19$. }
\label{fig:heavyquark}
\end{figure}

A more direct analysis of the fate of heavy quark bound states again
can come from the analysis of thermal meson correlation functions
constructed with heavy quarks \cite{Datta02,Umeda02}. These first,
exploratory studies show that pseudo-scalar ($\eta_c)$ and vector
($J/\psi$) correlation functions show almost no temperature dependence
across the phase transition and can survive in the high temperature
phase at least for temperatures $T\lsim 1.5 T_c$. 
In Fig.~\ref{fig:heavyquark} we show results for the vector meson
spectral function at temperatures below and above $T_c$. They are
compared with results for
screening masses which are indicated by horizontal lines in this figure.
As can be seen, the pole masses, identified as the location of the 
first peak in the spectral functions, coincide with the screening 
masses up to $T=1.5T_c$. 
Whether the apparent broadening shown in the figure and
reported in Ref.~\cite{Umeda02}
for pseudo-scalar and vector meson resonances is 
premature remains to be seen. 
At $3T_c$ on the other hand there is no 
evidence for a pole in the spectral function anymore and the 
threshold in the spectral function is unrelated to the value 
of the screening mass which itself moves away from its value
at the lower temperatures.
Thus there is evidence for modifications of spectral
properties at high temperature
and there are
also first hints that the vector meson
resonance is dissolved at least at $T\simeq 3 T_c$ \cite{Datta03}.
  
\section{Summary}

The survey of recent lattice calculations presented here shows that
there has been significant progress in our understanding of the
thermodynamics of strongly interacting matter since this field
has been reviewed for the last time in this series of
publications \cite{DeTar_qgp2}. We now have detailed quantitative 
information on the QCD equation of state at high temperature and 
the transition temperature to the plasma phase. Moreover, there
exist now first calculations for non-vanishing chemical potential
that allow us to judge the influence of a non-zero baryon number
density on the transition temperature and the equation of state.
Although these calculations are at present restricted to the region
of high temperature and small values of the chemical potential they
give first insight into the QCD phase diagram in the $T-\mu$ plane
and even gave first evidence for the presence of a critical point 
in this plane which has been anticipated in phenomenological models.
Nonetheless, lattice calculations still are performed with quark
masses larger than those realized in nature and we have to proceed
to smaller quark masses to provide reliable input to the analysis 
of current and future heavy ion experiments. 

One of the outstanding problems in QCD thermodynamics is to gain a 
quantitative understanding of thermal modifications of hadronic 
properties at high temperature. Although it is generally expected
that hadronic bound states cannot survive in the high temperature
plasma
phase we have little quantitative information about their dissolution 
in a hot thermal medium. Lattice calculations have provided many
indirect information on thermal modifications of hadron properties,
e.g. through the calculation of hadronic screening lengths and 
susceptibilities. However, it was only recently that we gained 
first insight into thermal modifications of hadron masses through
the application of statistical tools like the maximum entropy
methods. Hopefully, this will lead in the future to a direct 
verification of the disappearance of light and heavy quark bound states,
the structure of quasi-particle excitations in the plasma phase as 
well as experimentally and phenomenologically important observables like 
dilepton rates or even transport coefficients.

Progress in lattice studies of the QCD phase diagram as well as the
calculation of hadron properties have significantly been influenced 
through new conceptual ideas and the application of new calculational
schemes. Nonetheless, progress in lattice calculations has always 
been closely related to the progress made in computer technology.
Very soon a new generation of computers with a sustained speed in the
teraflops range will become widely available to the lattice community.  
It is to be expected that this will allow us to perform much more 
realistic numerical calculations with parameters that reflect the
quark mass spectrum realized in nature.

\section*{Acknowledgements}
This work has been supported by the DFG under grant FOR 339/2-1.

\begin{appendix}

\section{Appendix}

In this appendix we collect some formulae hopefully useful
in the context of our discussion of hadronic correlation
functions. Full account of 
the imaginary time formalism (at finite temperature)
is given in textbooks \cite{Muenster,LeBellac,Kapusta}. 

For the generic case of a spinless Bose field
the Feynman propagator in real, Minkowski time $t$ is given
by
\begin{equation}
D(t,\vec x;t',\vec x') =
\langle \, {\cal T} \{ \hat\phi(t,\vec x)
\hat\phi^\dagger(t',\vec x')\} \, \rangle  \quad .
\end{equation}
Here, the expectation value is taken at finite temperature,
\beq
\langle \, {\cal T}  \hat\phi(t,\vec x)
\hat\phi^\dagger(t',\vec x') \, \rangle
= \frac{1}{Z} \, {\rm tr } \,
 {\cal T} \hat\phi(t,\vec x)
\hat\phi^\dagger(t',\vec x')
e^{-\hat H / T} \quad ,
\label{eq:trace}
\eeq
with $\hat H$ being the Hamiltonian and $Z$
the partition function.
The Feynman propagator can be written as the sum of
two terms
\beq
D(t,\vec x;t',\vec x') =
\Theta(t-t') D^> (t,\vec x;t',\vec x') +
\Theta(t'-t) D^< (t,\vec x;t',\vec x')
\eeq
where
\beq
D^> (t,\vec x;t',\vec x') =
\langle \,  \hat\phi(t,\vec x)
\hat\phi^\dagger(t',\vec x') \, \rangle =
D^< (t',\vec x';t,\vec x) \quad .
\label{eq:dgt_def}
\eeq
The cyclicity of the trace in Eq.~\ref{eq:trace} leads to the KMS condition,
\beq
D^> (t;t') = D^< (t+i/T;t') \quad .
\label{eq:KMS}
\eeq 
In fact, the functions $D, D^>$ and $D^<$ only depend
on coordinate differences because of translational
invariance. Thus, their Fourier transformation is
given as, for instance,
\beq
D^> (t-t',\vec x-\vec x') =
\int \frac{d^3 \vec p}{(2 \pi)^3}
\int_{-\infty}^{+\infty} \frac{d p_0}{2 \pi}
e^{-i p_0 (t-t') + i \vec p (\vec x - \vec x')}
D^> (p_0,\vec p)~ .
\label{eq:dgt_fourier}
\eeq
The spectral density $\sigma(p_0,\vec p)$ is defined as
\beq
\sigma(p_0,\vec p) = 2 \pi \left[ D^>(p_0,\vec p)
 - D^<(p_0,\vec p) \right] \quad .
\label{eq:sigma_def}
\eeq
This also shows that $\sigma(p_0,\vec p)$ is an odd function
of the energy variable $p_0$.
Note that the spectral density not only depends on the
Lorentz invariant $p^2$ as at zero temperature but also
on the scalar product $np$ with $n_\mu=(1,\vec 0)$ which
arises from the presence of the heat bath.
By means of Eqs.~\ref{eq:dgt_def} and \ref{eq:dgt_fourier}
it is also obtained as 
\beq
\sigma(p_0,\vec p) = 2 \pi
\int d^3 \vec x
\int_{-\infty}^{+\infty} d t \,
e^{ i (p_0 t-\vec p \vec x )} \,
\langle \, [ \hat\phi(t,\vec x) ,
\hat\phi^\dagger(0,\vec 0) ] \, \rangle  \quad .
\eeq
Eq.~\ref{eq:sigma_def}, together with the KMS relation
can be inverted to give
\beq
D^>(p_0,\vec p) =
\frac{1}{1-e^{-p_0/T}} \frac{\sigma(p_0,\vec p)}{2 \pi} \quad. 
\label{eq:dgt_sol}
\eeq

In imaginary time, $t=-i \tau$ with $\tau$ real,
one can introduce a ``time'' ordering in $\tau$ and
define
\beq
\Delta(\tau,\vec x;\tau', \vec x') =
\langle T \{ \hat\phi(\tau,\vec x)
\hat\phi^\dagger(\tau',\vec x')\} \rangle \quad .
\eeq
The periodicity condition now follows from Eq.~\ref{eq:KMS}
\beq
\Delta(\tau,\vec x) \equiv \Delta(\tau,\vec x;0,\vec 0) =
\Delta(\tau-1/T,\vec x;0,\vec 0) \quad .
\eeq
As a consequence the Fourier spectrum consists
of discrete Matsubara frequencies,
\beq
\omega_n = 2 \pi n T, n = 0, \pm 1, \pm 2, ... \quad ,
\label{eq:boson_mats}
\eeq
at which the Fourier transform is defined
\beq
\Delta(i \omega_n, \vec p) =
\int d^3 \vec x
\int_{0}^{1/T} d \tau \,
e^{ i \omega_n \tau- i \vec p \vec x } \,
\Delta(\tau,\vec x;0,\vec 0) \quad .
\label{eq:ftransform}
\eeq
The inverse Fourier transform then includes a sum over
Matsubara frequencies
\beq
\Delta(\tau,\vec x)  =
T \sum_n \int \frac{d^3 \vec p}{(2 \pi)^3}
e^{ -i \omega_n \tau + i \vec p \vec x } \,
\Delta(i \omega_n, \vec p) \quad .
\eeq
Due to the identification
\beq
\Delta(\tau) = D^> (-i \tau)
\label{eq:iftransform}
\eeq
in the $\tau$ interval $[0,1/T]$ one then obtains from
Eqs.~\ref{eq:ftransform} and \ref{eq:dgt_sol} 
\beq
\Delta(i \omega_n, \vec p) =
\int_{-\infty}^{+\infty} d p_0
\frac{\sigma(p_0,\vec p)}{p_0-i \omega_n}  \quad .
\label{eq:delta_spec}
\eeq
Under certain assumptions, this expression can
be analytically continued to complex frequencies
$z$. In fact, for values $z=k_0 \pm i \epsilon$
with $k_0$ real, one obtains the retarded and
advanced propagators,
\begin{eqnarray}
D_R(t) & = &
\Theta(t) \, \langle \, \hat\phi(t) \hat\phi^\dagger(0)
\, \rangle \quad , \nonumber \\
D_A(t) & = & -
\Theta(-t) \, \langle \, \hat\phi(t) \hat\phi^\dagger(0)
\, \rangle \quad ,
\end{eqnarray}
as
\beq
D_{R/A} (k_0) = \mp \Delta(k_0 \pm i \epsilon) \quad .
\eeq

On the lattice one usually studies the temporal
correlator at fixed momentum $\vec p$.
By doing the sum over Matsubara frequencies this
can easily be obtained from Eqs.~\ref{eq:iftransform} and
\ref{eq:delta_spec} as
\begin{eqnarray}
G_T(\tau, \vec p) &=& \int d^3 x \; \Delta(\tau, \vec x) e^{- i \vec x \vec p}
\nonumber \\
&=&
\int_{-\infty}^{+\infty} d p_0 \;
\sigma(p_0, \vec p) \; e^{-\tau p_0} [\Theta(\tau) + n_B(p_0)]
\end{eqnarray}
with the Bose distribution
\beq
n_B(p_0) = [\exp(p_0/T) - 1]^{-1} \quad .
\eeq
Without loss of generality we take $\tau$ positive,
$\tau \epsilon [0,1/T]$, to arrive at Eqs.~\ref{eq:temp_corr},
\ref{eq:kernel}.

Instead of the temporal correlator, most (lattice) analyses have
been concentrating on the spatial correlation functions.
These depend of course on the same spectral density but
are different Fourier transforms of it
in $\tau$ and $\vec x _\perp = (x,y)$
\begin{eqnarray}
G (i \omega_n, \vec p _\perp, z) & = &
\int_0^{1/T} d \tau \int d^2 x_\perp \,
\exp[i \omega_n \tau - i \vec p _\perp \vec x _\perp]
\;\Delta(\tau, \vec x)  \nonumber \\
& = & \int_{-\infty}^{+\infty} \frac{d p_z}{2 \pi} \, e^{i p_z z} \,
\int_{0}^{+\infty} d p_0
\frac{ 2 p_0\; \sigma(p_0, \vec p _\perp, p_z) }{p_0^2 + \omega_n^2}
\label{eq:spatial}
\end{eqnarray}
Projecting onto vanishing transverse momentum
and vanishing Matsubara frequency then gives Eq.~\ref{eq:spatial0}.

For a free stable boson of mass $M$ the spectral density
has been given by Eq.~\ref{eq:free_boson}.
Setting the matrix elements to 1,
in Euclidean momentum space the propagator is thus obtained as
\beq
\Delta(i \omega_n, \vec p) = \frac{1}{\omega_n^2 + \omega_p^2}
\eeq
Computing the correlator in imaginary time gives
\begin{eqnarray}
G_T( \tau, \vec p) & = & \frac{1}{2 \omega_p}
\frac{e^{-\omega_p \tau} + e^{-\omega_p (1/T-\tau)}}
{1-e^{-\omega_p/T}}
\nonumber \\
& = & \frac{1}{2 \omega_p} K(\omega_p,\tau)
\end{eqnarray}
where
\beq
\omega_p = \sqrt{\vec p^{\,2} + M^2}
\eeq
In the limit of vanishing momentum the correlator thus
decreases with the mass modulo periodicity, 
Eq.~\ref{eq:free_temp}.
On the other hand, the spatial correlator is computed as
\beq
G (i \omega_n, \vec p _\perp, z)  =
\frac{1}{2 \omega_{\rm sc}} e^{-\omega_{\rm sc} z}
\eeq
where $\omega_{\rm sc}$ is given as
\beq
\omega_{\rm sc} = \sqrt{\omega_n^2 + \vec p_\perp ^{\,2} + M^2}
\eeq

The opposite limit is reached for two freely propagating
quarks contributing to the spectral density.
Here the starting point is the momentum space representation
of the meson correlation functions.
To leading order perturbation theory
one has to evaluate the self-energy diagram shown in Fig.~\ref{fig:bubble}a
in which the internal quark lines represent a bare quark propagator
$S_F(i \omega_n, \vec p)$ \cite{Friman}. The quark propagator can be expressed 
in terms of its spectral function\footnote{We use the convention as in
Ref.~\cite{LeBellac} where 
$\gamma_4 = i \gamma_0$ and $p_0 \rightarrow -i p_4 = i \omega_n$.}
$\sigma_{\rm free} (p_0, \vec{p})$,
\beq
\sigma_{\rm free}(p_0, \vec p) =  \epsilon(p_0) \;
(\psla+m) \; \delta(p_0^2 - \vec p^{\, 2} - m^2)
\eeq
and is conveniently written as
\beq
S_F(i \omega_n, \vec{p}) = 
\int_{-\infty}^{\infty}
 d p_0
\frac{\sigma_{\rm free} (p_0, \vec{p})}{p_0 - i \omega_n}
\label{eq:freequarkprop}
\eeq
Contrary to Eq.~\ref{eq:boson_mats} the fermion Matsubara frequencies are
odd integer multiples of $\pi T$,
\beq
\omega_n = (2n+1) \pi T 
\label{eq:fermion_mats}
\eeq
A free thermal meson spectral function is then obtained from the
imaginary part of the correlation function in momentum space,
confer Eq.~\ref{eq:delta_spec},
\begin{eqnarray}
\Delta_H(i \omega_m, \vec p) &=& N_c T \sum_n \int
\frac{d^3 \vec k}{(2 \pi)^3} {\rm tr} \left[
\Gamma_H S_F(i \omega_n, \vec k) \Gamma_H^\dagger
S_F^\dagger(i \omega_m - i \omega_n, \vec p - \vec k) 
\right]  \nonumber \\
& &
\label{eq:freequarkprop_2}
\end{eqnarray}
where $\omega_m$ denotes a boson frequency and $\omega_n$
a fermion one. The matrices $\Gamma_H$ ensure projection
on the spin and parity quantum numbers of a mesonic
state $H$.

The above analysis for the free thermal quark-antiquark correlators
can be extended to the leading order HTL approximation.
In this way
important medium effects of the quark-gluon plasma such as effective
quark masses and Landau damping are taken into account.
The HTL-resummed fermion propagator is obtained from
Eq.~\ref{eq:freequarkprop}
by replacing the free spectral function $\sigma_{\rm free}$
with the HTL-resummed
spectral function which for massless quarks
is given by \cite{Braaten,Kli82}
\begin{equation}
\sigma_{\rm HTL} (p_0, \vec{p}) =
 \frac{1}{2} \sigma_{+} (p_0,p)(\gamma_0 - \hat{p}\, \vec{\gamma} )
+\frac{1}{2} \sigma_{-} (p_0,p)(\gamma_0 + \hat{p}\, \vec{\gamma} )
\label{htl_sigma}
\end{equation}
with $\hat{p} = \vec{p}/p$, $p=|\vec{p}|$, and
\begin{eqnarray}
\sigma_{\pm} (p_0, p) & = &
   \frac{p_0^2 - p^2}{2 m_T^2}
   [\delta (p_0 - \omega_{\pm}) + \delta (p_0 + \omega_{\mp})]
   +\beta_{\pm} (p_0, p) \Theta(p^2 -p_0^2)
\nonumber \\
\beta_{\pm} (p_0, p) & = &
   \frac{m_T^2}{2}
   \left(p \mp p_0 \right)\left\{
\left[ p(p \mp p_0) + m_T^2 \left( 1 \pm \frac{ p \mp p_0}{2p}
\ln\frac{p+p_0}{p-p_0} \right)\right]^2\right. \nonumber \\
& & \hspace*{2cm} \left.+ \Bigl[ {\pi \over 2}
m_T^2 \frac{p \mp p_0}{p} \Bigr]^2
\right\}^{-1}
\label{htl_spec}
\end{eqnarray}
Here $\omega_{\pm} (p)$ denote the two dispersion relations of quarks
in a thermal medium \cite{Braaten,Kli82} and $m_T=g(T)T/\sqrt{6}$ is the
thermal quark mass. In addition to the appearance of two
branches in the thermal quark dispersion relation
the HTL-resummed
fermion propagator also receives a cut-contribution below the light-cone
($p_0^2<p^2$), which results from
interactions of the valence quarks with gluons in the thermal medium
(Landau damping).
Furthermore, an explicit temperature dependence only
enters through $m_T(T)$.
It also should be noted that the HTL resummed quark propagator
is chirally symmetric in spite of the appearance of an effective quark
mass \cite{Kli82}.

\end{appendix}

\end{document}